\documentclass[journal]{IEEEtran}
\pdfoutput=1
\usepackage[english]{babel}
\usepackage{float}
\usepackage{amsfonts}
\usepackage{amssymb}
\usepackage{amsthm}
\usepackage{booktabs}
\RequirePackage{CJKnumb}
\newtheorem{myDef}{Definition}
\newtheorem{proposition}{Proposition}

\newtheorem{lemma}{Lemma}
\newtheorem{assumption}{Assumption}
\usepackage{graphicx}
\usepackage{subfigure}
\usepackage{amssymb}
\usepackage{amsmath}
\usepackage{amssymb}
\usepackage{amsfonts} %% 这里是调用一些相关的宏包
\usepackage{amsthm}
\usepackage{amsfonts}
\usepackage{algorithm}
\usepackage{algorithmic}
\usepackage{mathrsfs}

\usepackage{braket}
\usepackage{booktabs}
\usepackage{bm}
\usepackage{bbm}

\usepackage{caption}
\usepackage{CJK}
\usepackage{color}

\makeatletter
\newcommand{\blue}[1]{{\color{black}{#1}}}

\newcommand{\m}[1]{{\color{black}{#1}}}
\newcommand{\bk}[1]{{\color{black}{#1}}}
\newcommand{\rj}[1]{{\color{black}{#1}}}
\newcommand{\taes}[1]{{\color{black}{#1}}}
\newcommand{\mktaes}[1]{{\color{black}{#1}}}
\newcommand{\mkktaes}[1]{{\color{black}{#1}}}
\newcommand{\iotj}[1]{{\color{black}{#1}}}
\newcommand{\chuan}[1]{{\color{black}{#1}}}
\newcommand{\Riotj}[1]{{\color{black}{#1}}}
\newcommand{\aciotj}[1]{{\color{black}{#1}}}
\makeatother

\usepackage{cases}
\usepackage{cite}

\usepackage{epsf}
\usepackage{epsfig}

\usepackage{graphicx}
\usepackage{gensymb}

\usepackage{indentfirst}

\usepackage{multirow}
\usepackage{makecell}

\usepackage{subfigure}
\usepackage{setspace}

\usepackage{textcomp}

\usepackage{url}
\usepackage[justification=centering]{caption}%标题居中  usepackage{caption2}

\allowdisplaybreaks

\allowdisplaybreaks

\usepackage{lineno} 
\begin{document}
% \begin{CJK}{UTF8}{gbsn}
    
% \pagewiselinenumbers % 按页重新编号 
% \switchlinenumbers	 % 双栏，单栏删除该行

\title{\blue{Stochastic Geometry-Based Performance Evaluation for LEO
Satellite-Assisted Space Caching}}

\author{

\IEEEauthorblockN{Chunyi Ma$^{1}$,  
                    Jiajie Xu$^{2}$, ~\IEEEmembership{Member,~IEEE,}
                   Jianhua Yang$^{1}$,
                   Mustafa A. Kishk$^{3}$, ~\IEEEmembership{Member,~IEEE}
}
\thanks{
\IEEEauthorblockA{$^{1}$  School of Automation, Northwestern Polytechnical University, Xi'an 710129, China},
\IEEEauthorblockA{$^{2}$ Computer, Electrical, and Mathematical Science and Engineering Division, King Abdullah University of Science and Technology, Thuwal, Saudi Arabia},
\IEEEauthorblockA{$^{3}$  Department of Electronic Engineering, Maynooth University, Maynooth, Ireland},
(e-mail: machunyi@mail.nwpu.edu.cn, jiajie.xu.1@kaust.edu.sa, yangjianhua@nwpu.edu.cn, mustafa.kishk@mu.ie.).
This work is supported by the short-term visiting program of Northwestern Polytechnical University. \blue{(Chunyi Ma and Jiajie Xu are co-first authors.)}
}	
}

% \graphicspath{{figure/}}

\maketitle

% \tableofcontents  % 生成目录
\begin{abstract}
\taes{\iotj{To achieve the Internet of Things (IoT) vision}, Mobile Edge Computing (MEC) is a promising \mktaes{technology} aimed at providing low-latency computing services to \Riotj{user equipment (UE)}. \mktaes{However, \Riotj{terrestrial MEC network} \Riotj{struggles} to provide service to \iotj{UEs} in \Riotj{remote and maritime region}.}} Low Earth Orbit (LEO) satellite networks have the potential to overcome geographical restrictions and provide seamless global coverage for \iotj{UEs}. In this paper, \chuan{we provide the first attempt to \Riotj{use stochastic geometry to} investigate the performance of implementing space caching with LEO satellites (SATs) \Riotj{in the MEC network}.} We \taes{\chuan{study} a LEO satellite-assisted space caching MEC network}, and \taes{LEO SATs can be equipped with servers to \taes{enable} space caching, \mktaes{with the advantage of seamless coverage to assist terrestrial CSs \Riotj{for serving UEs in remote or maritime reigon}}.} \taes{\mktaes{Using} stochastic geometry and queuing theory, we establish an analytical framework for this MEC network.} \mktaes{Meanwhile,} we develop association strategies for \iotj{UEs} to connect with LEO SATs or CSs and utilize stochastic geometry to derive uplink and downlink coverage probabilities, \rj{considering the diversity of task and service types}. \rj{On this basis,} we employ the queuing theory to \rj{calculate the average delay to evaluate the system performance.} Through Monte Carlo simulations and \mkktaes{numerical results}, \taes{the system performance is \mkktaes{evaluated}. \Riotj{The results show the potential of SAT spatial caching in improving the performance of the MEC network.} \mktaes{Additionally, our results reveal useful insights such as the} significant impact \mktaes{of the altitude and number of LEO SATs on the} average delay of the network, providing \mkktaes{helpful system-level recommendations for the design and configuration} of the space-caching MEC network.}

% Your abstract.
\end{abstract}

\IEEEpeerreviewmaketitle

% \vspace{-1mm}
\section{Introduction}
% \vspace{-1mm}
\subsection{Motivation}
% \vspace{-1mm}
{W}{ith} \iotj{the rapid development of IoT technology,} \blue{\iotj{smartphones, tablets, mobile routers, and a large number of user devices accessed through the IoT are becoming increasingly popular.} \iotj{This trend has significantly contributed to the explosive growth of} emerging applications such as Augmented Reality (AR)/Virtual Reality (VR), video streaming, web browsing, social networking, and online gaming, \rj{thus expanding to take on more and more compute-intensive and latency-sensitive tasks.}} \rj{\chuan{Nevertheless}, \chuan{the constrained processing capacity of \iotj{user equipments (UEs)} leads} to the challenge of low-latency communication and computation requirements for \iotj{IoT} networks\chuan{, when providing communication connectivity to \iotj{UEs} with wireless access.} \chuan{To address this problem,}} several new technologies have emerged, \chuan{among which the }\rj{mobile edge computing (MEC) tecgnology, which aims to deploy some servers at the edge of the network closer to the \iotj{UE}s, has received increasing research attention. \chuan{Because} edge servers are cached with services and computational resources for tasks, and UEs can connect to them \iotj{to get services}. \chuan{In this way, MEC} can improve the \taes{task access and computation efficiency} of the \iotj{UEs}, \taes{providing intelligent and timely solutions for various tasks and applications, such as healthcare and autonomous driving \cite{wang2014cache, gu2021modeling}.}} \taes{Since edge servers have limited storage and computational resources but are beneficial to \chuan{reduce the load of central cloud servers CSs and reduce network latency, while CSs have abundant computational resources but increase the latency for serving} remote \iotj{UE}s, \chuan{it is necessary to integrate CSs and edge servers, in order to exploit the complementary characteristics of both servers, which is in fact a key network architecture to cater for delay requirements \cite{zhang2021computation}. Furthermore, in practice, since} the request tasks of \iotj{UE}s are usually of multiple types, the different types of services cached by edge servers should also be considered as part of the MEC network \mktaes{parameters}.}

\rj{In the MEC network, \iotj{the applications of UEs can be partitioned into} executable tasks that can be executed on servers with different service types. Specifically, \iotj{UE}s offload their tasks \m{through uplink to} MEC servers, \taes{the servers provide computing services for the \iotj{data contained in} the tasks, and the \iotj{UE}s download the processed data results} \m{through downlink} \cite{gu2021modeling}. \taes{Due to the extensive potential and interdisciplinary properties of MEC, MEC technologies typically involve wireless communications and mobile computing \cite{ko2018wireless}, so the success of edge computing operations is also affected by the coverage of the communication link.}} However, \iotj{the lack of} terrestrial network infrastructure \iotj{is a common problem} in many rural and remote regions. \Riotj{Especially in maritime areas, the terrestrial network infrastructure makes it difficult to provide coverage and services to the UEs.} \iotj{And with the growing popularity of IoT networks and applications, \chuan{this problem is exacerbated by} the more common presence of UEs in these \Riotj{areas}, \chuan{leading to a continuous growth in demand for services in these areas   \cite{cheng2019space}}. Thus,} the current terrestrial MEC network still suffers from insufficient coverage and poor performance globally \cite{lai2021cooperatively,castro2014remote}. \rj{\chuan{Particularly,} when terrestrial infrastructure is damaged due to unexpected situations such as natural disasters, terrestrial MEC networks are unable to provide adequate coverage and timely and effective application services to \iotj{UE}s.} In recent years, the emergence of the Low Earth Orbit (LEO) satellite (SAT) network has \chuan{provided} an effective way to solve the above challenges, and has attracted extensive attention from researchers. LEO SAT can overcome geographical limitations and achieve global three-dimensional seamless coverage, with communication delay ranging from a few milliseconds to several tens of milliseconds to meet the needs of most application scenarios. Especially with the unprecedented growth of data flow, LEO SATs need to provide \iotj{UE}s with seamless and uninterrupted access around the globe and support them with a variety of computing services. For example, computing tasks generated by \iotj{UE}s in remote areas and oceans, where terrestrial network communication infrastructures are unavailable, can only be processed through the LEO SAT networks \cite{10599330}. \taes{Moreover, \Riotj{with the rapid advancements in onboard processing technologies, the launch cost of LEO satellites is gradually decreasing, }and recent research has indicated that LEO SATs already have the potential to incorporate ``high-throughput communication components'' \cite{fenech2015high,zech2015lct} and ``mass storage components''\cite{huang2018envisioned, jia2017collaborative} using existing technologies \cite{lai2021cooperatively}. Consequently, it can be argued that LEO SATs \chuan{could carry the service and computing resources to serve UEs in MEC} as the space caching \Riotj{to become edge computing nodes, and this technology is becoming a reality}.} \Riotj{The use of LEO satellite space caching in MEC to provide services and computation for UE tasks, which facilitates the realization of seamless services for UE tasks in remote or maritime areas.} \taes{Moreover, one of the key research areas for future MEC network deployment is the direct processing and computing of tasks on LEO SATs} \cite{alsharoa2020improvement, cheng2019space, liu2020task, tang2021computation}. \taes{However, investigating the specific impact of SAT space caching on network performance in MEC networks while exploiting the advantages of SAT is \mktaes{missing} in current \mktaes{literature,} and it is difficult to find a tractable framework for analysing the performance of MEC networks with SAT space caching assistance.}

\taes{\mktaes{Based on the above discussion}, we consider a SAT-assisted MEC network and provide an analytical framework for \chuan{performance evaluation}. We investigate the potential of using LEO SATs in theory to implement space caching to better understand the impact of SAT space caching in MEC networks, which facilitates access to a wide range of computational resources and services for \iotj{UE}s on a global scale.} To the best of our knowledge, \blue{this is the first attempt to} utilize stochastic geometry tools to tractably model and analyze the performance of a space-caching MEC network consisting of terrestrial \iotj{UE}s, terrestrial CSs, and LEO SATs. \blue{A detailed description of the contributions of this work is provided in Sec. I-C. But first, in the following subsection, we discuss relevant literature.}

\vspace{-2mm}
\subsection{Related Work}
% \vspace{-1mm}
\taes{Over the past decade, stochastic geometry has become a standard tool for modeling and analyzing the performance of wireless networks, creating an active research area \cite{ko2018wireless}. For example, \Riotj{Poisson} point process (PPP), has been used to model node locations in various wireless networks, such as cellular networks \cite{andrews2016primer}, heterogeneous networks \cite{zhang2021computation}, cache-enabled networks \cite{yang2015analysis} and cognitive radio networks \cite{sakr2015cognitive}. \mktaes{In addition, stochastic geometry tools have been widely used to model, characterise and obtain design insights for wireless networks with randomly deployed nodes \cite{lu2021stochastic}. \chuan{In general, this tool can} capture the inherent stochastic characteristics of wireless networks, which are more \chuan{consistent in} actual network deployment scenarios \cite{elsawy2016modeling}. Consequently, it significantly enhances the accuracy and practicality of the model. Additionally, stochastic geometry provides a tractable analytical framework for studying and analysing network performance. Notably,} the power of modeling network systems using stochastic geometry is to allow network performance to be described by a relatively small set of functions of the network parameters. In \cite{ko2018wireless}, \iotj{to realize the vision of the IoT and smart cities}, the authors studied the network latency performance of a large-scale terrestrial MEC network under the constraints of Radio Access Network connectivity and Computer Server Network stability. The PPP distribution was used to model the terrestrial access nodes and computer servers, and the impact of network parameters on network delay performance was analysed in combination with queuing theory and parallel computing theory, which provided guidance for the planning and deployment of MEC networks. \mktaes{In \cite{gu2019modeling,gu2021modeling}, \iotj{as the IoT continued to evolve, it was driving the urgent need to improve the quality of service for user tasks,} the authors focused on the performance of a large-scale terrestrial stochastic MEC wireless network, where the tasks of users can be processed using local resources or offloaded to the MEC edge servers. Based on the coupling of communication and computation, combined with stochastic geometry and queuing theory, the network nodes were modeled using the PPP and the Poisson cluster process to derive the average interruption probability of task transmission. The Markov chain was used to model the task execution process, and queuing theory was used to synthesize the average latency of end-to-end task execution, thus analysing the impact of local computation in users on the network's performance on average latency.}}  \mktaes{The above studies have shown that stochastic geometry is an effective tool to study the performance of MEC networks, while queuing theory is commonly used to calculate the average delay. The average delay result is a commonly used performance metric to measure the performance of MEC networks. However, these terrestrial MEC networks are susceptible to geographic constraints that make it difficult to achieve full coverage of terrestrial access points.} 

\mktaes{Satellites can provide ubiquitous and stable network communication coverage to users worldwide with their seamless coverage. In particular, satellites are able to provide services to remote and maritime areas that lack terrestrial network infrastructure \cite{xu2023space}.} Consequently, the LEO SAT network has attracted much attention from academia and industry, and the satellite-terrestrial integrated network \blue{has the potential to become} dominant in the next few years \cite{saad2019vision, giordani2020non, jia2022hierarchical}. \aciotj{The stochastic geometry analytical framework simulates dynamic topologies by randomly modeling satellite positions \cite{al2021analytic}. Within the stochastic geometry framework, analytical expressions for LEO SAT network system performance metrics can be derived as functions of system parameters (e.g., number of satellites) \cite{okati2021modeling, al2022optimal}, making the system performance metrics highly tractable.} Due to the tractability of stochastic geometry, it has recently been extensively used to model and analyze LEO satellite constellations in the context of SAT-based communication networks \cite{okati2020downlink, talgat2020stochastic, talgat2020nearest, talgatanalysis2024, wang2024ultra}. \mktaes{The Binomial point process (BPP) is one of the most widely used for LEO SAT modeling \cite{talgat2020nearest}, and the network system-level performance obtained using BPP more closely aligns with the results of actual LEO SAT constellations \cite{wang2024ultra}.}

\Riotj{In order to provide computing services to remote areas}, \Riotj{the application and research of LEO satellite networks have become a research hotspot in MEC \cite{jia2021toward,tang2021computation}. In \cite{wei2024energy}, the authors proposed a joint caching optimization and user selection problem. To reduce network energy consumption, they leveraged unmanned aerial vehicle (UAV) caching to maximize the residual energy of the satellite. In \cite{he2024online}, the authors considered the constraints on node energy in space-air-ground integrated networks and proposed a time-slot capture method for dynamic networks, based on which they formulated a stochastic optimization problem as a way to maximize the average data offloading.} In \cite{cao2022edge}, \blue{authors} designed an optimized loading framework for LEO SAT edge-assisted multi-layer multi-access edge computing systems, which jointly optimized the allocation of communication and computing resources. \mktaes{The aim was to minimize system energy consumption while keeping lower computational delays.} \mktaes{In \cite{tang2021computation}, considering that LEO satellite networks can break through geographical constraints and achieve global wireless coverage, which would be an indispensable choice for future mobile communication systems, the authors proposed a hybrid cloud and edge computing LEO satellite network with a three-tier computational architecture, intending to enable terrestrial users to access computing services on a global scale. And they proposed a distributed algorithm to minimise the total energy consumption of terrestrial users.} \aciotj{The above studies have advanced the integration of SAT space caching in MEC networks and mainly focus on algorithmic research for optimizing MEC performance. However, these studies have not yet provided tractable analytical performance expressions to specifically analyze the impact of LEO satellites on MEC network performance. In some cases, the algorithms may need to be redesigned if the network parameters are adjusted. With the support of analytical results, a performance metric can be expressed as a function of parameters such as satellite altitude and number of satellites. Thus, the researchers can directly observe the algorithm's performance as the parameters change.}

% \vspace{-0.4mm}
Although stochastic geometry has proven to be a powerful tool for analyzing LEO SAT-enabled communication systems and revealing various useful insights, \aciotj{none of the aforementioned works on SAT} space caching has considered using stochastic geometry to investigate LEO satellite-based MEC networks. \m{LEO SATs and terrestrial networks involve different communication models and attenuation. There is no work that fully considers the impact of the special characteristics of LEO SATs on the performance of MEC networks,} which is the main scope of this work.

% \vspace{-4mm}
\subsection{Contribution}
% \vspace{-1mm}
In this paper, we propose \blue{an analytical framework for integrated terrestrial-satellite MEC networks}. We use stochastic geometry tools to model the network and analyze its performance, aiming to provide useful design guidance for the deployment of LEO SAT space caching. \rj{To create a more practical model, we use stochastic geometry to model the spatial distribution of the network nodes while considering several factors. These include multiple types of tasks and services as well as limitations of SAT caching and computational capacity. Additionally, the association strategy is designed so that the \iotj{UE} can select an appropriate connection, either to a LEO SAT or to a CS.} The main contributions of this paper are summarized as follows.

\begin{itemize}
    \item \blue{This work constitutes the first attempt to provide a tractable analytical framework using stochastic geometry tools for LEO satellite-assisted MEC networks}. We model the location of the LEO SATs using the BPP, and the terrestrial CSs and terrestrial \iotj{UE}s are modeled using the PPP. \taes{The system model takes into account the diversity of task types and service types, as well as the storage and computing capacity constraints of the LEO SATs.}

    \item \rj{An association strategy is established for the system that takes into account the diversity of task and service types, giving \iotj{UE}s the flexibility to choose the appropriate server to connect to. \taes{For different transmission links}, we solve the association probabilities for the LEO SAT link and the terrestrial CS link, respectively.} \mktaes{The association probability between the \iotj{UE}s and the servers at each tier can also be used to quantify the average load on the servers.}
    
     \item \m{Based on the coverage probability of different communication links and the association probability, we calculate the average delay using queuing theory to evaluate the system performance. According to the process of executing tasks in the MEC network, we calculate the average uplink transmission time, average response time, and average downlink transmission time for different task types.}

    \item \blue{\mktaes{The accuracy of the analytical results is verified by the Monte Carlo simulation. Furthermore,} we reveal useful insights related to the influence of} the number of SATs, altitude, and \iotj{UE} density parameters on the performance of the space-caching MEC network.
\end{itemize}

The rest of the paper is structured as follows. Section II describes the system model of the \blue{considered} space caching MEC network, including the channel model for uplink and downlink, the association strategy, and the service and computation models. Section III derives the coverage probability and association probability. Section IV provides an analytical expression for the average delay performance of the system. Section V presents Monte Carlo simulation and analysis results to analyze the network performance and provide system-level insights. Finally, section VI summarizes this paper.

\begin{table}[h]
\caption{Symbol description.}
\centering
\scalebox{0.95}{
\begin{tabular}{c|c}
\hline
\textbf{\taes{Notation}}         & \textbf{\taes{Description}}   \\ \hline
${p_s}$, ${p_c}$, ${p_u}$        & Transmit power of SAT, CS, \iotj{UE}   \\ \hline
$\sigma _S^2$, $\sigma _C^2$, $\sigma _U^2$   & Noise power at SAT, CS, \iotj{UE}          \\ \hline
$\mathrm{SR}( {\Omega,b_0,m})$ & Shadowed-Rician fading      \\ \hline
$\alpha $   & Path-loss exponent    \\ \hline
$N_s$   & The number of LEO SATs    \\ \hline
${\lambda _c}$, ${\lambda _u}$   & The density of CS, \iotj{UE}           \\ \hline
${f_s}$, ${f_c}$                & Frequency band at SAT, CS        \\ \hline
${\cal N}_s$    & Capacity of buffer at satellites   \\ \hline
$W$   & Bandwidth      \\ \hline
${\cal D}_i^{u}$      & Input data size of task         \\ \hline
${\cal D}_i^{d}$    & Output data size of task          \\ \hline
$F_k$   &  Computational capacity  of $k$-tier server \\ \hline
$\tau $                              & Threshold                             \\ \hline
$q_i$                                & The probability of generating task ${\cal T}_i$    \\ \hline
\end{tabular}
}
\label{Symbol Description}
\end{table}

% \vspace{-2 mm}
\section{System model}\label{system mode section}
\begin{figure}[h]
    \centering
    \includegraphics[width=0.45\textwidth]{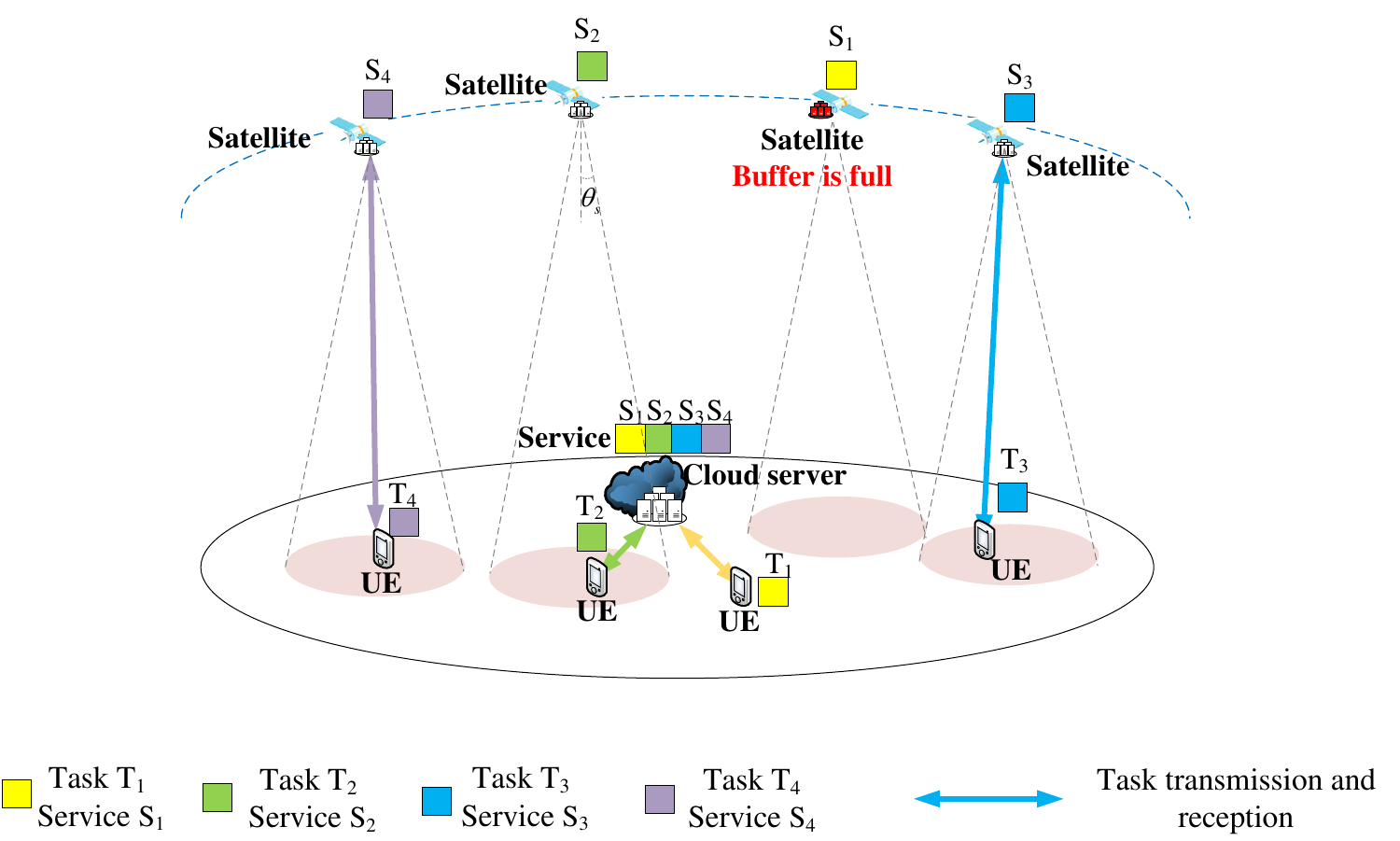}
    \caption{\Riotj{Schematic diagram of the network system.}}
    \label{system model}
\end{figure}

% \vspace{-2mm}
In this section, we \blue{describe} the network model, the channel model, the association strategy, and the service\blue{ and} computation model. \taes{They are the foundation of} analyzing the coverage probability and average delay. We consider a satellite-terrestrial integrated space caching \taes{MEC} network as shown in Fig. \ref{system model}. The network consists of the \taes{CSs, the \iotj{UE}s, and the LEO SATs}. \taes{Both CSs and LEO SATs have the capabilities of caching services and computing resources and a certain buffer space}. \taes{The space-caching SATs located at the edge of the network could \Riotj{serve for remote UEs and }relieve the load of the CSs. The \iotj{UE}s can offload computation-intensive and time-consuming tasks to the LEO SATs or CSs through uplink, and download task results through downlink.}  %(of this network?)

\vspace{-2.8mm}
\subsection{Network Model}
\vspace{-0.5mm}

% three consecutive processes: transmission, computation, and reception.By placing the \blue{cache copies} on the LEO SATs or CSs,

The locations of CS and \iotj{UE} located on the ground are modeled as two independent PPP distributions, denoted as ${\Phi_{c}}$ with density ${\lambda _{c}}$ and ${\Phi _u}$ with ${\lambda _u}$, respectively. \taes{PPP has been widely used for accurate modeling of terrestrial network nodes \cite{andrews2016primer,gu2021modeling}.} The locations of LEO SATs are modeled as BPP on a sphere concentric with the Earth, denoted as ${\Phi _s}$. \taes{BPP has been widely proven to be an effective distribution for modeling the node distribution of the LEO SAT network \cite{okati2020downlink, talgatanalysis2024, talgat2020stochastic, talgat2020nearest, wang2024ultra}.}\Riotj{\footnote{\Riotj{The Doppler frequency caused by the mobility of the LEO satellite can be estimated and mitigated by the mature pre-compensation method for UEs \cite{wang2019near}.}}} The altitude of the LEO SAT above the ground is ${a_s}$, and the radius of the sphere \blue{over} which the \blue{satellites are distributed} is ${r_s} = {r_e} + {a_s}$, where ${r_e}$ is the radius of the Earth. The number of LEO satellites is $N_s$, \rj{the density of LEO SAT is $\lambda_s$, $\lambda_s = {N_s}/{4\pi r_s^2}$}. 
% The LEO satellites have a beamwidth angle ${\theta _b}$, which provides 
% Illustration of the system model.
% As shown in Fig. \ref{system model}, CSs have all the task types requested
\vspace{-2.5mm}
\subsection{Channel model}
\taes{The channel link is responsible for transmitting the data of offloaded tasks from the \iotj{UE}s to the server, and after the server executes the task calculations, the \iotj{UE}s download the calculated results through the channel link.} Different types of servers use various channel models. Assuming that all \iotj{UE}s broadcast at the same fixed transmission power $p_u$, the constant transmission power of the LEO SATs is denoted as $p_s$, and that of the terrestrial CSs is denoted as $p_c$. In the rest of this paper, we will refer to the uplink between \blue{\iotj{UE} and SAT} as the U-S link and the downlink as the S-U link. Similarly, the uplink between \blue{\iotj{UE} and terrestrial CS is denoted as the U-C link, and the downlink as the C-U link. The channel fading for the ground CS follows Rayleigh fading and the path loss exponent is $\alpha$. The channel fading as for SAT in space follows Shadowed-Rician (SR) fading.}

\subsubsection{U-S Link}
The expressions for the received power at the serving SAT from the typical \iotj{UE} in the uplink is
\begin{equation}
    \rho _{}^{U - S} = p_u{L_u}{| {{h_0}} |^2},
\end{equation}
where ${| {{h_0}} |}^2$ represents the squared
Shadowed-Rician fading power. \aciotj{Existing research shows that the SAT link generally experiences free space attenuation \cite{vatalaro1995analysis} and the path loss exponent is most commonly used as 2 \cite{okati2022coverage}.} ${L_u}$ is the distance-dependent path loss function which is defined as\aciotj{\cite{talgatanalysis2024}}
\begin{equation}
\label{L path loss}
    {L_u}\left( {{D_\blue{u,U-S}}} \right) = {( {\frac{c}{{4\pi {f_s}{D_\blue{u,U-S}}}}} )^2},
\end{equation}
where $c$ is the speed of light, $f_s$ is the frequency, and $D_\blue{u,U-S}$ is the distance between typical \iotj{UE} and serving SAT in the uplink. $h_0$ represents the channel gain, which follows Shadowed-Rician fading (SR fading). According to \cite{alfano2007sum}\cite{xu2023space}, the cumulative distribution function (CDF) of the squared Shadowed-Rician fading power $|h_0|^2$ is as follows 
\begin{multline}
\setlength\abovedisplayskip{3pt}
\setlength\belowdisplayskip{3pt}
\label{SR fading}
{F_{|h_0|^2}}\left( t \right) =  {( {\frac{{2{b_0}m}}{{2{b_0}m + \Omega }}} )^m}{\sum\limits_{z = 0}^\infty \big[ {\frac{{{{\left( m \right)}_z}}}{{z!\Gamma \left( {z + 1} \right)}}( {\frac{\Omega }{{2{b_0}m + \Omega }}} )^z} }\\
 \times \Upsilon (z + 1,\frac{1}{{2{b_0}t}})\big],
\end{multline}
where, $\Gamma \left( \cdot  \right)$ denotes the Gamma function, $\Upsilon \left( { \cdot , \cdot } \right)$ is the lower incomplete gamma function, ${\left( m \right)_z}$ is the Pochhammer symbol, and $m$, $b_0$ and $\Omega $ are the parameters of the SR fading. We denote SR fading as $\mathrm{SR}\left( {\Omega ,{b_0},m} \right)$.

\subsubsection{U-C Link}

The received power at the serving CS from the typical \iotj{UE} in the uplink is modeled as
\begin{equation}
\rho^{U-C} = p_u \bk{(\frac{c}{4\pi f_cD_\blue{u,U-C}})^{ \alpha }}{g_0},
\end{equation}
where $\alpha $ is the path loss exponent, \bk{$f_c$ is the frequency of CS,} ${g_0}$ is the small-scale fading gain. Under Rayleigh fading, ${g_0}$ represent independent identically distributed (i.i.d.) exponential random \blue{variables} with $g_0 \sim \mathrm{exp}\left( 1 \right)$. \blue{$D_{u,U-C}$} is the random variable that represents the distance between typical \iotj{UE} and serving CS in uplink.

\subsubsection{S-U Link}
The expression for the received power at the typical \iotj{UE} from the nearest SAT in the downlink is
\begin{equation}
\rho^{S-U} = p_s{L_{d}}{\left| {{h_0}} \right|^2},
\end{equation}
where the expression for $L_d$ is related to distance $D_\blue{d,S-U}$ in downlink which is similar to (\ref{L path loss}). $h_0$ represents the channel fading, which follows SR fading as shown in (\ref{SR fading}).

\subsubsection{C-U Link}
The expression for the received power at the typical \iotj{UE} from the nearest CS in the downlink is
\begin{equation}
\rho ^{C-U} = p_c \bk{(\frac{c}{4\pi f_c D_\blue{d,C-U}})}^{ \alpha }{g_0},
\end{equation}
where $ D_\blue{d,C-U}$ is the random variable that represents the distance between serving CS and typical \iotj{UE} in the downlink. $g_0$ also follows the Rayleigh fading with $g_0 \sim \mathrm{exp}\left( 1 \right)$.

\vspace{-4mm}
\subsection{Association Strategy}
\vspace{-1.5mm}
\taes{As there are different types of servers (CSs and SATs) in the network, this heterogeneous network can become more complex in terms of association strategy compared to the traditional single station type network. We use the cell association strategy based on the maximum biased average power, i.e., the \iotj{UE} will associate with the most powerful server among the nearest servers (CS and SAT). The fading is averaged out in the cell association, i.e., the fading is ignored in the association strategy \cite{jo2012heterogeneous}.} \rj{\m{The bias factor mainly represents the tendency of \iotj{UE} to choose a particular server tier.}} \taes{SATs are expected to be designed to have a bias towards admitting \iotj{UE}s, thus reducing the pressure on CSs to connect to \iotj{UE}s. The presence of a bias factor would allow the \iotj{UE} to have a more flexible choice of associated server.} Let $D_c$ and $D_s$ represent the distance from the typical \iotj{UE} to the nearest CS and the nearest SAT, respectively. \taes{According to the association strategy, a typical \iotj{UE} will be associated with a server of tier $k \in \{c, s\}$ as expressed below}
% \rj{In association strategy, \taes{when considering the average power of associated servers,}}. A typical MU will associate with a server of \blue{tier $k \in \{c, s\}$,}
\begin{align}
    k=\begin{cases}
        s, &if  \quad {p_s}{B_s}{( {\frac{c}{{4\pi {f_s}{D_s}}}} )^2}  \geq {p_c}{B_{c}}\bk{(\frac{c}{4\pi f_c D_c})}^{ \alpha } \\
        c, &if \quad {p_c}{B_{c}}\bk{(\frac{c}{4\pi f_c D_c})}^{ \alpha } > {p_s}{B_s}{( {\frac{c}{{4\pi {f_s}{D_s}}}} )^2}
    \end{cases},
\end{align}
where $B_{c}$ represents the association bias factor for CS tier, $B_{s}$ represents the association bias factor for SAT tier. \Riotj{In this paper, the determination of the bias factor is considered from two main aspects. On the one hand, it aims to promote more UEs to associate with the satellite, to fully utilize the full coverage characteristics of the satellite, and to alleviate the load of CS. On the other hand, we use the bias factor to compensate for the effect of satellite altitude on satellite average power, which can make the UE associate with the satellite even if the average received power of the satellite is less than that of the CS.}
% $k \in \left\{ {c,s} \right\}$, if

% \begin{equation}
% {K} = \arg \max \left\{ {{p_c}{B_{c}}D_c^{ - \alpha },{p_s}{B_s}{{\left( {\frac{c}{{4\pi {f_s}{D_s}}}} \right)}^2}} \right\},
% \end{equation}

The association probability ${{\cal A}_k}$ of \iotj{UE} is
\begin{align}
\label{defin AP}
 {{\cal A}_k}
 % & =\mathbb{P}\blue{(K = k)}\nonumber\\
% & 
= \begin{cases}
\mathbb{P}\big({p_s}{B_s}{( {\frac{c}{{4\pi {f_s}{D_s}}}} )^2} \geq {p_c}{B_{c}}\bk{(\frac{c}{4\pi f_c D_c})}^{ \alpha }\big),   & k = s\\
\mathbb{P}\big({p_c}{B_{c}}\bk{(\frac{c}{4\pi f_c D_c})}^{ \alpha } > {p_s}{B_s}{( {\frac{c}{{4\pi {f_s}{D_s}}}} )^2}\big),  & k = c
\end{cases}.
\end{align}

\taes{\mktaes{A basic expression is given in \eqref{defin AP}} for the association probability. Since we are considering the situation where there are multiple tasks and service types in the network, we \mktaes{further stratify} the tier of the associated server, which we explain in detail in Sec. \ref{coverage analysis section}, and the results can be found in Sec. \ref{AP section}, \textbf{Lemma} \ref{lemma AP}. In particular, when there is task diversity, the association probability of the cell association strategy based on the maximum biased average power can quantify the average number of \iotj{UE}s associated with one server in each tier \cite{jo2012heterogeneous}. In addition, we assume that \iotj{UE}s supported by the same server cell use different orthogonal resource blocks, so there is no interference among \iotj{UE}s associating with the same server.} 

% \vspace{-3mm}
\subsection{Service and Computation Model}
% \vspace{-1mm}
Define a set of tasks in the system as $\left\{ {{{\cal T}_i}\left| {i = 1,2,...,{\cal M}} \right.} \right\}$, \rj{${\cal M}$ is the total number of tasks.} Each task ${{\cal T}_i}$ can be represented by a 4-tuple, i.e., ${{\cal T}_i} = \left( {{{\cal S}_i},{{\cal L}_i},{\cal D}_i^{u},{\cal D}_i^{d}} \right)$\blue{, where} ${{\cal S}_i}$ is the required service type for the execution of ${{\cal T}_i}$, ${{\cal L}_i}$ represents the necessary Central Processing Unit (CPU) cycles to accomplish ${{\cal T}_i}$, and ${\cal D}_i^\taes{u}$ as well as ${\cal D}_i^\taes{d}$ denote the input and output data sizes of   ${{\cal T}_i}$, respectively. \taes{It is well known that edge caching servers have limited storage and computing resources compared to CSs, where CSs have more resources \cite{lai2021cooperatively}.} It is assumed that a typical \iotj{UE} generates a task ${{\cal T}_i}$ with probability $q_i$, and each SAT only caches a single service ${{\cal S}_i}$, with probability $q_i$. CSs have services for all types of tasks. We refer to SAT caching ${{\cal S}_i}$ as type-$i$ SAT. We use $F_k$ to denote the computational capacity of the $k$-tier server, which is defined as the number of CPU cycles per second. According to queuing theory, we assume that the computation service process follows a first-in-first-served discipline. \rj{Since both SATs and \iotj{UE}s are movable, here we only consider the transmission and computational performance of the tasks within a certain time slot.\Riotj{\footnote{\Riotj{The LEO satellites can use solar panels to obtain energy\cite{yang2016towards, ali2021power}. We assume that the LEO SATs have enough energy to serve UEs during a time slot.}}}} For a type-$i$ task ${{\cal T}_i}$, the service rate of tier $k$ servers is given by $\mu _k^i = {F_k}/{{\cal L}_i}$. The service time for task ${\cal T}_i$ on tier $k$ servers follows the exponential distribution with $1/\mu _k^i$. We adopt the M/G/1 and M/M/1/N queuing models \cite{kleinrock1975queue} to analyze the computation service process for CS and SAT, respectively. 

\mktaes{Since we use the cell association strategy based on maximum biased average power, the mean association area of an associated server in the $k$ tier is $\frac{{{\cal A}_k}}{{{\lambda _k}}}, k \in \{s,c\}$ \cite{singh2013offloading}. Given that \iotj{UE}s follow the PPP distribution, the arrival process for offloading tasks to the computing server follows a Poisson process with a specific arrival rate. The task arrival rate at a SAT with service ${\cal S}_i$ is \cite{zhang2021computation}} 
% the MUs 
\begin{equation}
\label{Pofld_arr}
\Lambda _{s,i}^{} = \frac{{{{\cal A}_{s,i}}{\lambda _u}}}{{{\lambda _s}}},
\end{equation}
where ${{\cal A}_{s,i}}$ is the association probability that a typical \iotj{UE} with ${\cal T}_i$ is associated with a SAT.

Due to the limited computational capacity of SAT, \blue{we} assume that the computing buffer of each SAT is bounded by ${{\cal N}_s}$. The SAT \blue{does not} provide service to additional \iotj{UE}s when the computing buffer of SAT is full. As stated in \cite{zhang2021computation}, the probability that the buffer of type-$i$ SAT is not full is defined as
\begin{align}
% \scalebox{0.85}{$
P_{\rm{ofld}}^i &= 1 - P(\rm {The{\kern 1pt} {\kern 1pt} {\kern 1pt} {\kern 1pt} satellite's{\kern 1pt} {\kern 1pt} {\kern 1pt} {\kern 1pt} buffer{\kern 1pt} {\kern 1pt} {\kern 1pt} is{\kern 1pt} {\kern 1pt} {\kern 1pt} full} ) \nonumber\\
& = 1 - \frac{{\left( {1 - {\rho _i}} \right)\rho _i^{{\cal N}_s}}}{{1 - \rho _i^{{\cal N}_s + 1}}},
% $}
\label{Pofld}
\end{align}
where ${\rho _i} = \frac{{\Lambda _{s,i}}}{{{\mu _s^i}}}$, $\Lambda _{s,i}$ is the arrival rate given by (\ref{Pofld_arr}).

The CSs are supposed to be equipped with an abundant computational buffer. \taes{Since CS can serve all types of tasks, the arrival rate at a CS is}
\begin{equation}
\Lambda _c^{} = \sum\limits_{i = 1}^{\cal M} {\frac{{{{\cal A}_{c,i}}{\lambda _u}}}{{{\lambda _{c}}}}},
\end{equation}
where ${{\cal A}_{c,i}}$ is the probability that a typical \iotj{UE} with type-$i$ task associates \blue{with the} CS.

% \vspace{-3mm}
\section{Coverage Analysis}\label{coverage analysis section}
% \vspace{-1mm}
\taes{Since there are a total number of $\cal M$ tasks in the system}, all servers can be considered as ${\cal M} +1$ tiers, where CSs are denoted as tier 0, and SATs, which cache the service ${\cal S}_i$, are denoted as tier $i$ $(1  \leq i  \leq {\cal M})$. We denote the number of satellites in the tier $i$ as $N_{s_i}$, and the density of \iotj{UE}s with task ${\cal T}_i$ as ${\lambda _{{u_i}}}$, where $\sum_{i = 1}^{\cal M} {{N_{{s_i}}} = {N_s}} $, $\sum_{i = 1}^{\cal M} {{\lambda _{{u_i}}} = {\lambda _u}}$. The number of type-$i$ SATs available for offloading is denoted as $N_i$, ${N_i} = {N_{{s_i}}} \times P_{ofld}^i$, where $P_{ofld}^i$ is obtained according to (\ref{Pofld}). \rj{The system we consider is a multi-tier network that includes multiple tasks and service types. \Riotj{We assume that the nodes with different tasks and service types do not interfere with each other.} \taes{For the considered system, the time cost consists of transmission time and response time.}}

% \vspace{-4mm}
\subsection{Downlink Analysis}
% \vspace{-1.5mm}
Since the path loss and channel attenuation are different for both SAT and CS, the signal-to-Noise Ratio (SNR) for different scenarios at the typical \iotj{UE} needs to be analyzed separately. \Riotj{According to Slivnyak's theorem, arbitrarily selecting a UE as the typical UE to be analyzed does not affect the generality of the performance analysis results.}

For the S-U link, we note that the distance between the typical \iotj{UE} and the nearest \blue{type-$i$ offloadable} SAT is ${D_\blue{d,S-U}}$, which is a random variable. The SNR at the typical \iotj{UE} from the associated \blue{the type-$i$ offloadable} SAT is given by %serving Sat
\begin{equation}
% \begin{array}{l}
\mathrm{{SNR}}_{S_i}^D = \frac{\rho^{S-U}}{{\sigma _U^2}} = \frac{{p_s{{({\frac{c}{{4\pi {f_s}{\blue{D_{d,S-U}}}}}})}^2}{{\left| {{h_0}} \right|}^2}}}{{\sigma _U^2}},
% \end{array}
\end{equation}
where $c$ refers to the speed of light, ${f_s}$ is carrier frequency. ${h_0}$ is the SR fading, $\sigma_U^2$ is the noise power \blue{at the} \iotj{UE}.

For the C-U link, the typical \iotj{UE} connects to the nearest CS \blue{at a distance of ${D_{d,C-U}}$}. 
The SNR at the typical \iotj{UE} located at a distance \blue{${D_{d,C-U}}$} from the nearest CS is denoted as 
\begin{equation}
\mathrm{SNR}_C^D = \frac{\rho ^{C-U}}{{\sigma _U^2}}=\frac{p_c \bk{(\frac{c}{4\pi f_c D_{d,C-U}})}^{ \alpha }{g_0}}{{\sigma _U^2}}.
\end{equation}
% where $g_0$ is Rayleigh fading with $g_0 \sim \rm exp(1)$.

\begin{figure}[h]
    \centering   \includegraphics[width=0.25\textwidth]{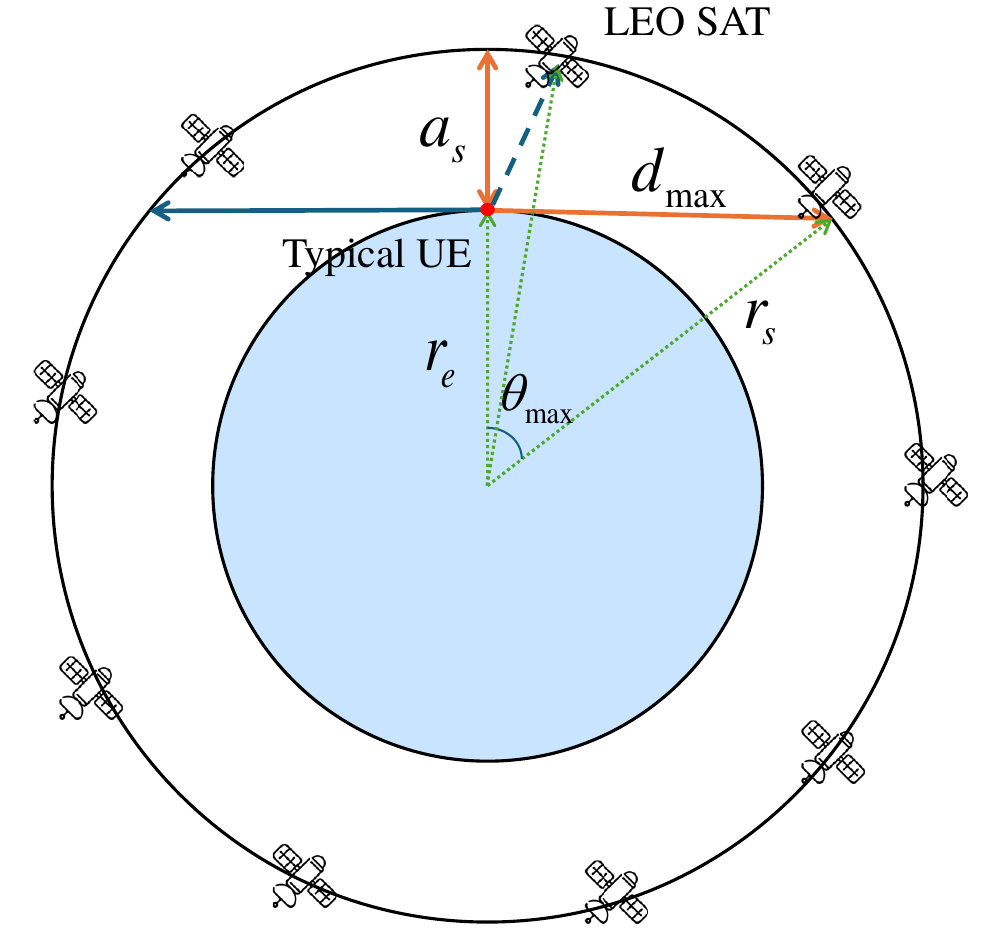}
    \caption{S-U downlink geometry diagram.}
    % Coverage Geometry of S-U link
    \label{SU system model}
\end{figure}

In order to derive the downlink coverage probability, we have to determine the distribution of the distance between typical \iotj{UE} and serving SAT or CS, so we first calculate the cumulative distribution function (CDF) and probability density function (PDF) of the distance. The detailed geometric relationship for the S-U link is shown in Fig. \ref{SU system model}.

\begin{lemma} \blue{Contact Distance Distribution of S-U link.
The CDF and PDF of the distance ${D_{d,S-U}}$ between  the nearest offloadable type-$i$ SAT and the typical \iotj{UE}  is \cite{talgat2020stochastic}}
\begin{align}
&\quad \blue{{F_{{D_{d,S-U}}}}}( {{x}} )= \nonumber\\
 &
\begin{cases}
0,& {x} < {a_s}\\
1 - {[ {1 - \frac{1}{\pi }\arccos ( {1 - \frac{{x^2 - a_s^2}}{{2{r_e}{r_s}}}} )} ]^{\blue{{N_i}}}},& {a_s} \le {x} < {d_{\max }}\\
1-{[1-\frac{1}{\pi}\arccos(\frac{r_e}{r_s})]^{N_i}},&  {d_{\max }} \le {x}
\end{cases},
\label{CDF S-U}
\end{align}
and the PDF of distance \blue{${D_{d,S-U}}$} is 
\begin{align}
&\quad{f_{\blue{{D_{d,S-U}}}}}(x)= \nonumber\\
&\begin{cases}
\frac{{{x}\blue{{N_i}}}}{{\pi {r_e}{r_s}}{\sqrt {1 - {( {1 - \frac{{x^2 - a_s^2}}{{2{r_e}{r_s}}}} )^2}} }}\\
\quad\quad   \times {\Big[ {1 \!-\! \frac{1}{\pi }\arccos \big( {1 \!-\! \frac{{x^2 - a_s^2}}{{2{r_e}{r_s}}}} \big)} \Big]^{\blue{{N_i}} - 1}}\!,& \!{a_s} \le {x} < {d_{\max }}\\
0,& else
\end{cases},
\label{S-U pdf}
\end{align}
where, ${d_{\max }}= \sqrt {2r_ea_s + a_s^2}$.
\label{cdf pdf SUlink}
\end{lemma}

The CDF and PDF of the distance \blue{${D_{d,C-U}}$} between typical \iotj{UE} and
the nearest CS \blue{can be easily obtained using the void probability of PPP as follows}
\begin{equation}   
\setlength\abovedisplayskip{3pt}
\setlength\belowdisplayskip{3pt}
{F_{\blue{{D_{d,C-U}}}}}( {\blue{x}} ) = 1 - \exp \left( { - {\lambda _c}\pi x^2} \right),
 \label{CDF C-U}
\end{equation}
and
\begin{equation}
\setlength\abovedisplayskip{3pt}
\setlength\belowdisplayskip{3pt}
\label{PDF C-Ueq}
{f_\blue{{D_{d,C-U}}}}( {\blue{x}} )  = 2{\lambda _c}\pi x^{}\exp \left( { - {\lambda _c}\pi x^2} \right),
\end{equation}
where $\lambda_c$ is the density of CS.
\label{lemma CUlink cdf pdf}

Applying the law of total probability, the downlink coverage probability for each task generated is obtained as the sum of the product of the association probability of each tier and the corresponding coverage probability. The downlink coverage probability of the system is
\begin{align}
P_{{\mathop{\rm cov}} }^D( \tau )&\! =\! \sum\limits_{i = 1}^{\cal M} {{q_i}\mathbb{P}({\mathrm{SNR}^D > \tau \left| {Typical{\kern 1pt} {\kern 1pt} {\kern 1pt} \iotj{UE}{\kern 1pt} {\kern 1pt} with  {\kern 1pt} {\kern 1pt} task{\kern 1pt} {\kern 1pt} {\kern 1pt} {{\cal T}_i}} \right.} )} \nonumber\\
&\! =\! \sum\limits_{i = 1}^{\cal M} {{q_i}\!\left[ \blue{\mathcal{A}_i}P_{{S_i}}^D\left( \tau  \right)\! +\! \blue{\mathcal{A}_0}P_{{C_i}}^D\left( \tau  \right) \right]},
\label{total dwonlink CP}
\end{align}
in which,
\begin{equation}
P_{{S_i}}^D( \tau ) = \mathbb{P} ( {\mathrm{SNR}_{{S_i}}^D > \tau} ),
\label{defin CP SU}
\end{equation}
and
\begin{equation}
P_{{C_i}}^D( \tau  ) = \mathbb{P}( {\mathrm{SNR}_{\blue{C}}^D > \tau \left| {\iotj{UE}{\kern 1pt} {\kern 1pt} with{\kern 1pt} {\kern 1pt} {\kern 1pt} {\kern 1pt} {{\cal T}_i}} \right.}).
\label{defin CP CU}
\end{equation}   

\blue{In order to solve (\ref{total dwonlink CP}), we need to respectively derive (\ref{defin CP SU}) (\ref{defin CP CU}), representing the downlink coverage probability between the typical \iotj{UE} with the type-$i$ offloadable SAT and the CS.} In the SAT-related communication link, the channel attenuation obeys the SR fading, and the CDF of the Shadowed-Rician fading squared power $|h_{0}|^2$ is given in \eqref{SR fading}. For the convenience of the solution, we use the approximation of the SR fading \taes{which is described next}.

\begin{proposition}
\label{pro 1}
The Shadowed-Rician fading, ${|h_0|^2}\sim \mathrm{SR}\left( {{\Omega} ,{b_0},m} \right)$ can be approximated as Gamma random variable $H \sim \Gamma \left( {{\alpha _s},{\beta _s}} \right)$, with shape and scale parameters ${\alpha _s}$ and ${\beta _s}$, respectively, such that
\begin{align}
{F_{|h_{0}|^2}}\left( t \right) = \frac{1}{{\Gamma \left( {{\alpha _s}} \right)}}\Upsilon ({\alpha _s},\frac{t}{{{\beta _s}}}),
% $}
\end{align}
where ${\alpha _s} = \frac{{m{{\left( {2{b_0} + \Omega } \right)}^2}}}{{4mb_0^2 + 4m{b_0}\Omega  + {\Omega ^2}}}$, ${\beta _s} = \frac{{4mb_0^2 + 4m{b_0}\Omega  + {\Omega ^2}}}{{m\left( {2{b_0} + \Omega } \right)}}$.

\taes{For the relative accuracy of this approximation, please refer to \cite{sellathurai2016user}\cite{abdi2003new}.}
\end{proposition}

In order to get a tractable form of the expression of coverage probability, we use an expression for a tight lower/upper bound for the CDF of a Gamma random variable $H \sim \Gamma \left( {{\alpha _s},{\beta _s}} \right)$ stated in \cite{alzer1997some}, \taes{and ${|h_0|^2}\approx H$}, the CDF of Gamma distribution can be tightly bounded as %\cite{talgatanalysis2024}\cite{thornburg2016performance}
\begin{align}
\label{app 2}
\setlength\abovedisplayskip{3pt}
 \setlength\belowdisplayskip{3pt}
 \begin{cases}
{F_H}( t ) \le {(1 - {e^{ - \frac{{\mu t}}{{{\beta _s}}}}})^{{\alpha _s}}}, & if \  {\alpha _s} \le 1\\
{F_H}(t) > {(1 - {e^{ - \frac{{\mu t}}{{{\beta _s}}}}})^{{\alpha _s}}},& if \   {\alpha _s} > 1
\end{cases},
\end{align}
where $\mu  = {\left( {{\alpha _s}!} \right)^{ - \frac{1}{{{\alpha _s}}}}}$.

\taes{This tight bound of \eqref{app 2} has been demonstrated to be an acceptable approximation to the CDF of the Gamma random variable, further details \mktaes{can be found in} \cite{jia2021uplink}. In \cite{talgatanalysis2024}, authors have employed this approximation to conduct derivation and simulation on LEO SAT coverage probability.}
\begin{lemma}
The downlink coverage probability 
 $P_{{S_i}}^D$  of S-U link is
 \begin{align}
% \begin{equation}
% \begin{aligned}
& P_{{S_i}}^D( \tau ) = \frac{{\blue{N_{i}}}}{{\pi {r_e}{r_s}}}\int_{{a_s}}^{{d_{\max }}} \Big[ {\sum\limits_{j = 1}^{{\alpha _s}} {\big( {_j^{{\alpha _s}}} \big)} {{( { - 1} )}^{j + 1}}\exp ( - \frac{{j\mu A x^2}}{{{\beta _s}}})} \Big]\nonumber\\
&\times\!\frac{{{x}}}{{\sqrt {1 \!-\! {( {1 \!-\! \frac{{x^2 - a_s^2}}{{2{r_e}{r_s}}}})^2}} }}{\Big[ {1 \!-\! \frac{1}{\pi }\arccos ( {1 \!-\! \frac{{x^2 - a_s^2}}{{2{r_e}{r_s}}}} )} \Big]^{\blue{N_{i}} - 1}} \mathrm{d}{x},
\label{S-U downlink CP}
\end{align}
where $\mu  = {\left( {{\alpha _s!}} \right)^{ - \frac{1}{{{\alpha _s}}}}}$, $A = \frac{{\tau \sigma _U^2}}{{\blue{p_s}( {\frac{c}{{4\pi {f_s}}}} )^2}}$.
% , ${N_s} = \sum\limits_{i = 1}^{\cal M} {{N_{{s_i}}}} $.
\label{lemma SU CP}
\end{lemma}
\emph{proof}: see Appendix A. 
 $\hfill\blacksquare$

The coverage probability $P_{{C_i}}^D$ of C-U downlink is
\begin{align}
&P_{{C_i}}^D\left( \tau  \right) = \nonumber\\
&2\pi {\lambda _c}\int_0^\infty  {\exp ( { - \tau \sigma _U^2{{( {{p^{-1}_c}})}}\bk{(\frac{c}{4\pi f_c})^{-\alpha}}x^\alpha })\exp \left( { - {\lambda _c}\pi x^2} \right)x\mathrm{d}{x}},
\label{C-U downlink CP}
\end{align}
where $\sigma _U^2$ is the noise variance of the \iotj{UE}, $\lambda_c$ is the density of CS.

The analytical expression (\ref{total dwonlink CP}) for the total downlink coverage probability of the system can be obtained by bringing (\ref{S-U downlink CP}), (\ref{C-U downlink CP}) to (\ref{total dwonlink CP}).
% \vspace{-1mm}
% \vspace{-2mm}
\subsection{Association Probability}\label{AP section}
Based on the cell association strategy with maximum biased average power, we solve for the association probability, which needs to be solved separately for SAT and CS because of the different channel models. \blue{As we previously stated, we utilize ${\cal A}_i$ to represent the \iotj{UE} with task  ${\cal T}_i$ associated with the nearest tier-$i$ offloadable SAT, and ${\cal A}_0$ to represent the \iotj{UE} with task ${\cal T}_i$ associated with the nearest CS.}

\begin{lemma} 
\label{lemma AP}
The association probability ${{\cal A}_k}$ is 
\begin{footnotesize}
% \begin{scriptsize}
\begin{align}
&{{\cal A}_k} = \nonumber\\
&\begin{cases}
{\displaystyle\int_0^{( {{a_s}Q_s^\blue{\frac{\alpha}{2}} })^{\frac{2}{\alpha }}} } {2\pi {\lambda _c}x \exp ({ - \pi {\lambda _c}{x^2}})\mathrm{d}x}+ 2\pi {\lambda _c}  \\
\quad \times \displaystyle\int_{( {{d_{\max }}{Q_s^\blue{\frac{\alpha}{2}}}} )^{\frac{2}{\alpha }}}^\infty{{{\big[ {1 \!-\! \frac{1}{\pi }\arccos ( {\frac{{{r_e}}}{{{r_s}}}} )}\big]}^{\blue{N_i}}} x\exp ( { - \pi {\lambda _c}{x^2}})\mathrm{d}x }\\
\quad +2\pi {\lambda _c}\displaystyle\int_{{{( {{a_s}Q_s^\blue{\frac{\alpha}{2}} })}^{\frac{2}{\alpha }}}}^{{{( {{d_{\max }}Q_s^\blue{\frac{\alpha}{2}}} )}^{\frac{2}{\alpha }}}} {{\!\!\big[ {1 \!-\! \frac{1}{\pi }\arccos ( {1 \! -\!\frac{{Q_s^{ - \alpha }{x^\alpha } \! -\! a_s^2}}{{2{r_e}{r_s}}}} )} \big]^{\blue{N_i}}}} x\\
\quad \times \exp ( { - \pi {\lambda _c}{x^2}} )\mathrm{d}x, &k = 0\\
\vspace{0mm}\\
 \frac{{\blue{N_i}}}{{\pi {r_e}{r_s}}}\displaystyle\int_{{a_s}}^{{d_{\max }}} {\exp ( { - {\lambda _c}\pi Q_s^2x^{\frac{4}{\alpha }}} )} \frac{x}{{\sqrt {1\! -\!{( {1 \! -\! \frac{{{x^2} - a_s^2}}{{2{r_e}{r_s}}}} )^2}} }}\\
\quad \times{\big[ {1  \! -\! \frac{1}{\pi }\arccos( {1 \! -\!\frac{{{x^2} - a_s^2}}{{2{r_e}{r_s}}}} )} \big]^{\blue{N_i} \!-\! 1}}\mathrm{d}x, &k = i
\end{cases},
\end{align}
% \end{scriptsize}
\end{footnotesize}
where ${Q_s} = {( {\frac{{{p_c}{B_c}}}{{{p_s}{B_s}}}} )^{\frac{1}{\alpha }}}{( {\frac{{4\pi {f_s}}}{c}})^{\frac{2}{\alpha }}}{\bk{(\frac{c}{4\pi f_c })}}$, ${d_{\max }} = \sqrt {2{r_e}{a_s} + a_s^2} $.
\end{lemma}
\emph{proof}: see Appendix B.
 $\hfill\blacksquare$
 
\subsection{Uplink Analysis}

\Riotj{For the U-S link, each SAT has a set of {UE}s within its {line of sight}. That means for a serving satellite, the typical UE is a random UE within its line of sight.} We use the random variable \blue{$D_{u,U-S}$} to represent the distance between the typical \iotj{UE} and the serving SAT. The SNR at the serving SAT from the typical \iotj{UE} is 
\begin{align}
\mathrm{SNR}_{\blue{S_i}}^U = \frac{{\rho ^{U-S}}}{{\sigma _S^2}} = \frac{{p_u{{( {\frac{c}{{4\pi {f_s}D_\blue{u,U-S}}}} )}^2}{{| {{h_0}}|}^2}}}{{\sigma _S^2}},
\end{align}
where, ${\sigma _S^2}$ is the noise power \blue{at the} SAT.

For U-C link, \blue{t}he SNR at the serving CS from the typical \iotj{UE} is
\begin{equation}
\mathrm{SNR}_C^U = \frac{{\rho^{U-C}}}{{\sigma _C^2}} = \frac{p_u \bk{{( {\frac{c}{{4\pi {f_c}D_\blue{u,U-C}}}} )}^{ \alpha }}{g_0}}{{\sigma _C^2}},
\end{equation}
where ${\sigma _C^2}$ is the noise power \blue{at the} CS.

\begin{figure}[H]
    \centering
\includegraphics[width=0.25\textwidth]{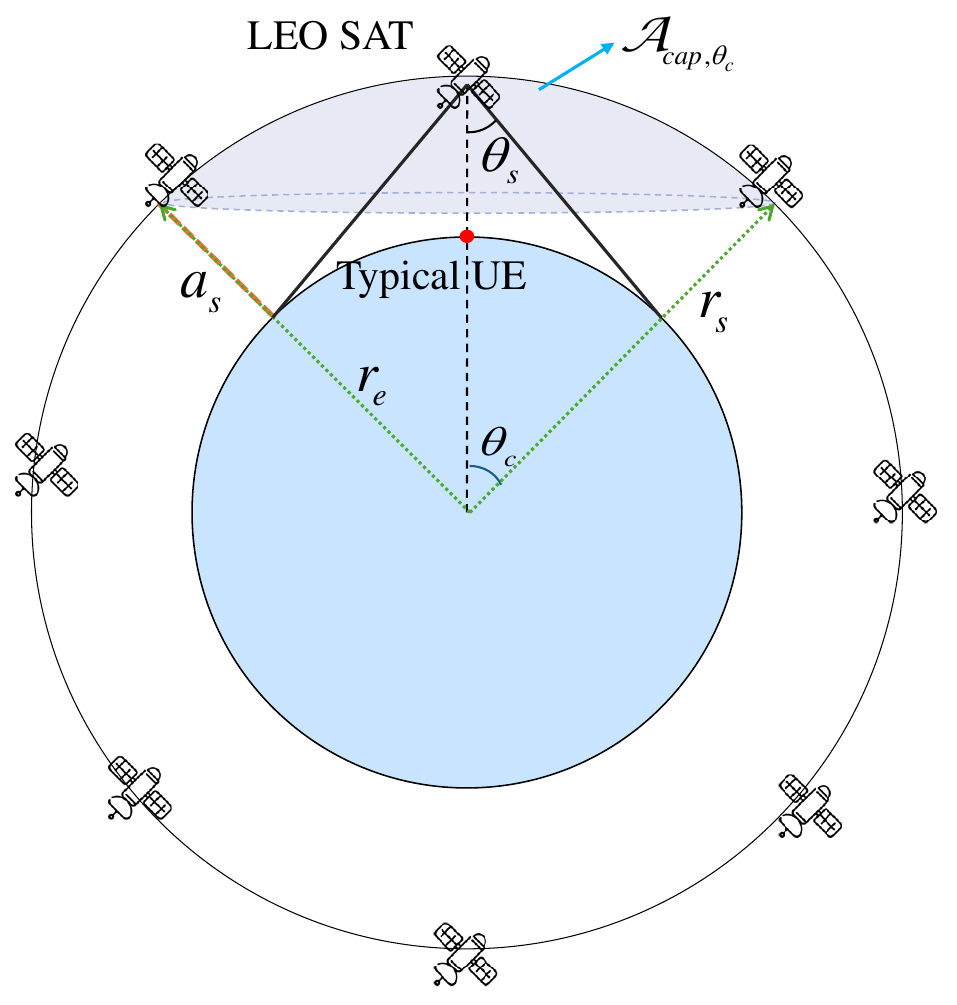}
    \caption{U-S uplink geometry diagram.}
    \label{USsystemmode}
\end{figure}

In order to calculate the uplink coverage probability, we need to solve the contact distance distribution of U-S link and U-C link. The specific geometric relationship for U-S link is shown in Fig. \ref{USsystemmode}.
\begin{lemma} Contact distance distribution of U-S link. The CDF and PDF of the distance $D_\blue{u,U-S}$ between typical \iotj{UE} and the tagged SAT is \cite{talgatanalysis2024}
\begin{align}
\label{US-cdf}
\scalebox{0.9}{$
{F_{D_\blue{u,U-S}}}( {\blue{x}}) = \begin{cases}
0, &{x} < {a_s}\\
\frac{1}{{{\theta _c}}}\arccos \left( {\frac{{r_e^2 + r_s^2 - x^2}}{{2{r_e}{r_s}}}} \right),& {a_s} \le {x} < {d_{\max }}( {{\theta _c}} )\\
1,&{d_{\max }}( {{\theta _c}} ) \le {x}
\end{cases},
$}
\end{align}
and
\begin{align}
\label{US-pdf}
\scalebox{0.9}{$
{f_{{D_\blue{u,U-S}}}}( {\blue{x}} ) = \begin{cases}
\frac{1}{{{\theta _c}{r_e}{r_s}}} \frac{{{x}}}{{\sqrt {1 - {{( {\frac{{r_e^2 + r_s^2 - x^2}}{{2{r_e}{r_s}}}})}^2}} }},& {a_s} \le {x} < {d_{\max }}( {{\theta _c}} )\\
0, & else
\end{cases}.
$}
\end{align}
where \m{${d_{\max }}\left( {{\theta _c}} \right) = \sqrt {r_s^2 - r_e^2 } $, $\theta _c =  \arccos (\frac{{{r_e}}}{{{r_s}}}) $} \blue{b}ased on the geometric relationship in Fig. \ref{USsystemmode}.

\end{lemma}

% \vspace{-3mm}

For the U-C link, the CDF and PDF of the distance $D_\blue{u,U-C}$ between serving CS and \iotj{UE} are \cite{andrews2016primer} 
\begin{equation}
    \begin{array}{l}
{F_{{D_\blue{u,U-C}}}}\left( {\blue{x}} \right) = 1 - \exp \left( { - {\lambda _c}\pi x^2} \right),
\label{U-C cdf}
\end{array}
\end{equation}
and
\begin{equation}
   \begin{array}{l}
{f_{{D_\blue{u,U-C}}}}\left( {\blue{x}} \right) = 2{\lambda _c}\pi x^{}\exp \left( { - {\lambda _c}\pi x^2} \right).
\end{array}
\label{U-C pdf}
\end{equation}

The uplink coverage probability of this system is given by	
\begin{align}
P_{{\mathop{\rm cov}} }^U( \tau ) &\! =\! \sum\limits_{i = 1}^{\cal M} {{q_i}\mathbb{P}( {\mathrm{SNR}^U} > \tau \left| {Typical {\kern 1pt} {\kern 1pt} {\kern 1pt} UE {\kern 1pt} {\kern 1pt} {\kern 1pt} with {\kern 1pt} {\kern 1pt} {\kern 1pt} {\kern 1pt} task{\kern 1pt} {\kern 1pt} {\kern 1pt} {{\cal T}_i}} \right.} )\nonumber \\
&\! =\! \sum\limits_{i = 1}^{\cal M} {{q_i}{\big[ \blue{{\cal A}_i} P_{{S_i}}^U\left( \tau  \right) + \blue{{\cal A}_0} P_{{C_i}}^U( \tau)} \big]},
\label{total uplink CP}
\end{align}
in which, 
\begin{equation}
P_{{S_i}}^U( \tau ) = w\mathbb{P}( {\mathrm{SNR}_{{S_i}}^U > \tau  } ),\\
\label{upCP U-S}
\end{equation}
where $w = 1 \!- \! {( {\frac{{1 + \cos {\theta _c}}}{2}} )^{{N_\blue{i}} \!- \! 1}}$. \m{$\theta_c$ is obtained by SAT line of sight, and}
\begin{align}
P_{{C_i}}^U(\tau) = \mathbb{P}( {\mathrm{SNR}_{\blue{C}}^U > \tau \left| \blue{\iotj{UE}{\kern 1pt} {\kern 1pt} with{\kern 1pt} {\kern 1pt} {\kern 1pt} {\kern 1pt} {{\cal T}_i}} \right.} ).
\label{upCP U-C}
\end{align}

In order to solve \eqref{total uplink CP}, we need to derive the uplink coverage probability of \iotj{UE} with SAT or CS, respectively, and the expression for solving \eqref{upCP U-S} \eqref{upCP U-C} is given directly below.
% [?]
\begin{lemma}
The coverage probability $P_{{S_i}}^U(\tau )$ of U-S uplink is 
\begin{align}
&P_{{S_i}}^U( \tau ) =\frac{w}{{{\theta _c}{r_e}{r_s}}}\int_{{a_s}}^{{d_{\max }}( {{\theta _c}} )} \!\!{\Big[ {\sum\limits_{j = 1}^{{\alpha _s}} {( {_j^{{\alpha _s}}})} {{( { \!-\! 1} )}^{j + 1}}\! \exp ( \!-\! \frac{{j\mu Ax^2}}{{{\beta _s}}}\!)}\!\Big]}\nonumber\\
&\qquad\qquad\qquad\qquad\qquad\quad\times\frac{{{x}}}{{\sqrt {1 - {{({\frac{{r_e^2 + r_s^2 - x^2}}{{2{r_e}{r_s}}}} )}^2}} }} \mathrm{d}\blue{x},
\end{align}
where $A = \frac{{\tau \sigma _S^2}}{{\blue{p_u}( {\frac{c}{{4\pi {f_s}}}} )^2}}$, $w = 1 \!- \!{\left( {\frac{{1 + \cos {\theta _c}}}{2}} \right)^{{N_\blue{i}} - 1}}$.
\label{lemma Pu-s CP}
\end{lemma}
\emph{proof}: see Appendix C.
 $\hfill\blacksquare$

The U-C uplink coverage probability is widely used in \cite{andrews2016primer} as follow
\begin{align}
&P_{{C_i}}^U( \tau ) = \nonumber \\
& 2\pi {\lambda _c}\int_0^\infty  {\exp ( { - \tau \sigma _C^2p_u^{ - 1}\bk{{(\frac{c}{4 \pi f_c})^{-\alpha}}}x^{\alpha } })\exp ( { - {\lambda _c}\pi x^2} )x\mathrm{d}\blue{x}}. 
\label{lemma U-C CP}
\end{align}

% \vspace{-1mm}
% \vspace{-3mm}
\section{Average delay Analysis}
In this section, we provide the average delay analysis for \blue{the considered system}. \taes{According to the service and computation process of the MEC network for the tasks of the \iotj{UE}s, clearly,} the average delay includes the average transmission time and the average response time. We first compute the uplink and downlink average transmission time expressions for task offloading and \mktaes{computation results} downloading based on the coverage probability. Then we derive the arrival rates at the \taes{type-$i$} SAT and the CS to obtain the average response time \taes{for task ${\cal T}_i$}.

\taes{We use an Orthogonal Resource Allocation (ORA) \cite{jo2012heterogeneous} scheme to allocate bandwidth, that is, for any server, the same bandwidth is allocated in a round-robin manner to the \iotj{UE}s served (i.e., the server load). We consider the total bandwidth is $W$. The bandwidth allocated to the typical \iotj{UE} is determined by the load of the server. The analysis of the load is given below.}

\begin{assumption}
Mean Load Approximation. For analytical simplicity, we assume that the load (the number of \iotj{UE}s associated with a server providing the service) is equal to its mean value \cite{singh2013offloading}.
\end{assumption}

The average number of \iotj{UE}s with task ${\cal T}_i$ connected to a SAT with service ${\cal S}_i$ \blue{is approximated as follows} \cite{singh2013offloading}
\begin{align}
\mathbb{E}[ {N_{{s_i}}^u} ] = 1 + \frac{{1.28{\lambda _{{u_i}}}{{\cal A}_{{s_i}}}}}{{{\lambda _{{s_i}}}}},
\end{align}
where ${{\cal A}_{{s_i}}}$ is the association probability that a \iotj{UE} connects a SAT with ${\cal S}_i$, ${\lambda _{{s_i}}} = {{{N_{s_i}}}}/{{4\pi r_s^2}}$.

The average number of \iotj{UE}s with task ${\cal T}_i$ connected to a CS is
\begin{align}
\mathbb{E}[ {N_{{c}}^u} ] = \frac{\lambda_u {\cal A}_{0i}}{\lambda_c},    
\end{align}
\blue{where \blue{$ {\cal A}_{0i}$} is the association probability that a \iotj{UE} with task ${\cal T}_i$ connects a CS.}

The bandwidth allocated to a typical \iotj{UE} is as follows
\begin{align}
\begin{cases}
% \left\{ \begin{array}{l}
W_c = \frac{W}{{N_c^u}}, &  \iotj{UE}{\kern 1pt} {\kern 1pt} connected{\kern 1pt} {\kern 1pt} {\kern 1pt} to{\kern 1pt} {\kern 1pt} {\kern 1pt} CS\\
W_{{s_i}}^{} = \frac{W}{{N_{{s_i}}^u}},& { \iotj{UE}{\kern 1pt} {\kern 1pt} connected{\kern 1pt} {\kern 1pt} {\kern 1pt} to{\kern 1pt} {\kern 1pt} {\kern 1pt} type-}i{\kern 1pt} {\kern 1pt} {\kern 1pt} \rm SAT
% \end{array} \right.
\end{cases}.
\end{align}

\taes{Given $\tau$, the average downlink transmission rate of the \iotj{UE} with task ${\cal T}_i$ connected to the CS is $\mathbb{E}[{W_c}{\log _2}(\tau  + 1)\mathbb{P}({\rm SNR}_C^D > \tau )] = W_c{\log _2}(\tau  + 1)P_{{C_i}}^D(\tau )$, similarly, the average uplink transmission rate is ${W_c}{\log _2}(\tau  + 1)P_{{C_i}}^U(\tau )$. The average downlink transmission rate of the \iotj{UE} with task ${\cal T}_i$ connected to the type-$i$ offloadable SAT is $\mathbb{E}[{W_{{s_i}}}{\log _2}(\tau  + 1)\mathbb{P}({\rm SNR}_{{S_i}}^D > \tau )] = {W_{{s_i}}}{\log _2}(\tau  + 1)P_{{S_i}}^D(\tau )$, similarly, the average uplink transmission rate is ${W_{{s_i}}}{\log _2}(\tau  + 1)P_{{S_i}}^U(\tau )$}.
% $\mathbb{1}( \cdot )$ represents the indicator function

For the task ${\cal T}_i$ \taes{with input data size ${\cal D}_i^u$}, the average uplink transmission time is
\begin{align}
\begin{cases}
T_{{c_i}}^{t,u}\left( \tau  \right) = \frac{{{\cal D}_i^{u}}}{{P_{{C_i}}^U( \tau )W_c{{\log }_2}\left( {\tau  + 1} \right)}},& task \ {\cal T}_i \  \iotj{UE}\!-\!CS \\
T_{{s_i}}^{t,u}\left( \tau  \right) = \frac{{{\cal D}_i^{u}}}{{P_{{S_i}}^U\left( \tau  \right)W_{{s_i}}^{}{{\log }_2}\left( {\tau  + 1} \right)}},&  task \ {\cal T}_i \ \iotj{UE}\!-\!SAT
\end{cases},
\end{align}
where, $ P_{{S_i}}^U( \tau )$ and $ P_{{C_i}}^U( \tau )$ are shown in Lemma \ref{lemma Pu-s CP} and (\ref{lemma U-C CP}).

For the task ${\cal T}_i$ \taes{with output data size ${\cal D}_i^d$}, the average downlink transmission time is
\begin{align}
\begin{cases}
T_{{c_i}}^{t,d}\left( \tau  \right) = \frac{{{\cal D}_i^{d}}}{{P_{{C_i}}^D\left( \tau  \right)W_c^{}{{\log }_2}\left( {\tau  + 1} \right)}},&task \ {\cal T}_i \  \iotj{UE}\!-\!CS\\
T_{{s_i}}^{t,d}\left( \tau  \right) = \frac{{{\cal D}_i^{d}}}{{ P_{{S_i}}^D\left( \tau  \right)W_{{s_i}}^{}{{\log }_2}\left( {\tau  + 1} \right)}},&  task \ {\cal T}_i \ \iotj{UE}\!-\!SAT
\end{cases},
\end{align}
where $P_{{S_i}}^D$ and $P_{{C_i}}^D$ are shown in Lemma \ref{lemma SU CP} and (\ref{C-U downlink CP}).

\begin{myDef}
Response time. The response time is the time spent by a task to be computed and processed on the server, which includes the waiting time and the service time. We use $T_{{s_i}}^r$ and $T_{{c_i}}^r$ to denote the response time of SAT and CS for the task ${\cal T}_i$, respectively.
\end{myDef}

% For task ${\cal T}_i$, the arrival rates at Sat and CS are
Based on the M/M/1/N queuing model used by SAT, the arrival rate of task ${\cal T}_i$ at the SAT is
\begin{equation}
{\Lambda _{s,i}} = \frac{{{{\cal A}_{{s_i}}}{\lambda _u}{q_i}P_{{S_i}}^U\left( \tau  \right)}}{{{\lambda _{s_i}}}},
\end{equation}

Based on the M/G/1 queuing model applied at CS, the arrival rate of task ${\cal T}_i$ at the CS is
\begin{equation}
{\Lambda _{c,i}} = \frac{{{{\cal A}_{0i}}{\lambda _u}{q_i}P_{{C_i}}^U\left( \tau  \right)}}{{{\lambda _c}}}.
\end{equation}

According to the Pollaczek-Khinchin mean formula \cite{kleinrock1975queue}, the average response time of task ${\cal T}_i$ at the CS can be determined as
\begin{equation}
T_{{c_i}}^r = \frac{1}{{\mu _c^i}} + \frac{{( {\sum\limits_{j = 1}^{\cal M} {{\Lambda _{c,j}}} }) \times ( {\sum\limits_{j = 1}^{\cal M} {\frac{{2{{\hat \Lambda }_{c,j}}}}{{{{\left( {\mu _c^j} \right)}^2}}}} } )}}{{2\big[ {1 - ( {\sum\limits_{j = 1}^{\cal M} {{\Lambda _{c,j}}} } ) \times ( {\sum\limits_{j = 1}^{\cal M} {\frac{{{{\hat \Lambda }_{c,j}}}}{{\mu _c^j}}} } )} \big]}}, 
\end{equation}
where ${\hat \Lambda _{c,i}} = \frac{{{\Lambda _{c,i}}}}{{\sum_{j = 1}^{\cal M} {{\Lambda _{c,j}}} }}$.

With reference to \cite{kleinrock1975queue}, for the type-$i$ SAT, the average response time for task ${\cal T}_i$ is
\begin{equation}
T_{{s_i}}^r = \frac{{{{\overline Q }_i}}}{{{\Lambda _{s,i}}\left( {1 - {\sigma _i}} \right)}},
\end{equation}
where ${\overline Q _i} = \frac{{{\rho _i}\left( {1 - \left( {{{\cal N}_s} + 1} \right)\rho _i^{{\cal N}_s} + {{\cal N}_s}\rho _i^{{\cal N}_s}} \right)}}{{\left( {1 - \rho } \right)\left( {1 - \rho _i^{{{\cal N}_s} + 1}} \right)}}$, ${\sigma _i} = \frac{{\rho _i^{{\cal N}_s}}}{{\sum_{n = 0}^{{\cal N}_s} {\rho _i^n} }}$, ${\rho _i} = \frac{{{\Lambda _{s,i}}}}{{\mu _s^i}}$.

In conclusion, the average delay of the typical \iotj{UE} with task ${\cal T}_i$ is
\begin{align}
T_{avg}^i\left( \tau  \right)& = \taes{\mathbb{E}[T_{respon.}+T_{trans.}]}\nonumber\\
&={{\cal A}_{{0i}}}\left[ {T_{{c_i}}^r + T_{{c_i}}^{t,u}\left( \tau  \right)+ T_{{c_i}}^{t,d}\left( \tau  \right)} \right]\nonumber\\
& \quad+ {{\cal A}_{{s_i}}}\left[ {T_{{s_i}}^r + T_{{s_i}}^{t,u}\left( \tau  \right) + T_{{s_i}}^{t,d}\left( \tau  \right)} \right].
\end{align}

% \vspace{-3mm}
\section{Numerical and Simulation Results}
% \vspace{-1mm}
% In this section, we analyze and validate the performance of the analyzed network using numerical analysis and Monte Carlo simulation. 
In this section, we analyze the system performance based on numerical results and validate the results using Monte Carlo simulation. We investigate the effect of different parameters on system performance. \taes{Simulation and analysis are performed on MATLAB R2020b.} \blue{The values of the simulation} parameters are shown in Table. \ref{value of parameters}. Unless otherwise stated, the parameter values for the simulation follow those in the table, most of the parameter values used are 
consistent with literature\Riotj{\cite{talgatanalysis2024, cao2022edge, xu2023space, zhang2021computation,xie2024computation, luglio2022performance}}. Other parameters and values are mentioned explicitly when used.

\begin{table}
\centering
\caption{The value of parameters.}
\scalebox{0.85}{
\begin{tabular}{c|c||c|c}
\hline
\textbf{Parameter} & \textbf{Value}       & \textbf{Parameter} & \textbf{Value} \\ \hline
${p_u}$       & $23$ dBm   &  ${\cal D}_i^{u}$      &  $ 0.5$ kb   \\ \hline
${p_c}$      & $45$ dBm     & ${\cal D}_i^{d}$      & $0.3$ kb      \\ \hline
${p_s}$      & $60$ dBm     & $r_e$    & $6371$ km  \\ \hline
$f_s$  & $2$ GHz &  $\lambda_u$  & $45\   \mathrm{points}/\mathrm{km}^2$\\ \hline
$\tau $  &  $0$ dB  &$\lambda_c$   &  $1\ \mathrm{point}/\mathrm{km}^2$\\ \hline
 $F_s$ &  {$3$} GHz  &  $F_c$  &  {$10$} GHz
  \\ \hline
$\sigma _U^2$ & $-98$ dBm    & ${\cal N}_s$    &   $2$     \\ \hline
$\sigma _C^2$ & $-117$ dBm   &  $\mathrm{SR}( {\Omega ,{b_0},m})$      & $\mathrm{SR}({1.29,0.158,19.4})$   \\ \hline
$\sigma _S^2$ & $-174$ dBm & $q_i$ & $0.25$ \\ \hline
$\alpha $ & $2.7$ & $W$  &  {$500$} MHz  \\ \hline
\m{$f_c$} & $1$ GHz & $Bs/Bc$ & $200$ \\   \hline
\end{tabular}
}
\label{value of parameters}
\end{table}
% \end{footnotesize}
First, we analyze the impact of SAT altitude and the number of SATs on the performance of the considered network system. The results of the variation of the average delay with the number of SATs are compared for different SAT altitudes of 500 km, 800 km, and 1000 km, respectively, with all other parameters held constant. 
% The Monte Carlo count is 3000.

% \vspace{-2mm}
\begin{figure}
    \centering   \includegraphics[width=0.4\textwidth]{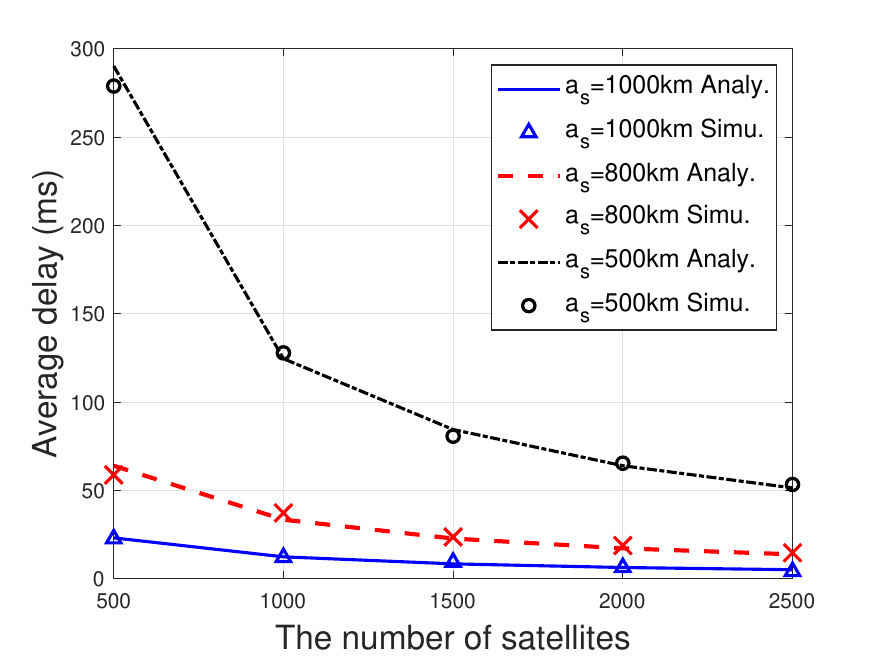}
    \caption{Effect of $a_s$.}
    \label{Effect of $a_s$}
\end{figure}

Fig. \ref{Effect of $a_s$} shows the impact of satellite altitude on the performance of the considered network system analyzed through analysis and simulation results. As the number of SATs increases, the average delay at different SAT altitudes shows a decreasing trend. The solid line represents the analytical results, and the marked point is the simulation results. \mktaes{The analytical and simulation results are matched to verify the correctness  and \aciotj{accuracy} of the theoretical analysis.} Overall, among the three altitudes, the higher the SAT altitude, the lower the average delay and the better the system performance. The lower the SAT altitude, the more significant the drop in average delay when the number of SATs is small, as can be seen from the extent of the decrease in the black line in Fig. \ref{Effect of $a_s$}. Eventually, for larger numbers of satellites, the average delays at different altitudes reach their own minimums and remain essentially flat, and the gap among their average delays is very small, but it is still the case that higher altitudes have the lowest average delays. This is explained by the fact that increasing the SAT altitude can enhance the coverage probability between the SAT and the \iotj{UE} to some extent, which has the effect of improving the system performance.

% \vspace{-2mm}
\begin{figure}
    \centering   \includegraphics[width=0.4\textwidth]{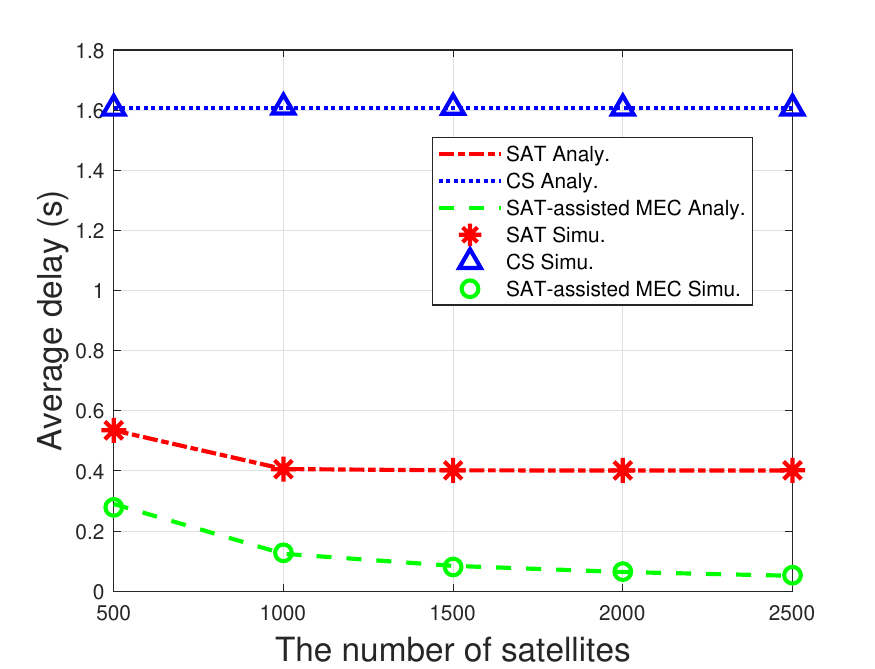}
\caption{\Riotj{Comparison of average delay of our system versus that of the individual server network.}}
    \label{Comparison of average delay of system and average delay of single server.}
\end{figure}

Fig. \ref{Comparison of average delay of system and average delay of single server.} shows the comparison between the average delay of the \blue{considered system} and the average delay when only the SAT or only the CS is operational, with the SAT altitude of 500 km. \blue{It is evident from the shown results that the considered system, where SATs and ground CSs are integrated, outperforms the scenarios that rely only on SATs or only on ground CSs from the perspective of average delay.} \aciotj{The average delay of the MEC network using only SAT is less than that of using only CS, which illustrates the effectiveness of SAT space caching. Meanwhile, the average delay of the SAT-assisted MEC network is lower than that of only CS, which demonstrates that the average delay performance is improved with the assistance of SAT space caching.} \Riotj{This phenomenon is analyzed as follows. The LEO satellite-assisted MEC network studied in this paper combines the strong service capability of CSs and the advantage of seamless coverage of satellite edge nodes. The existence of satellite edge nodes shares part of the load of CSs and enhances the service performance of CS. In addition, LEO satellites enhance the service capability of the MEC network to the remote UEs. Therefore, the SAT satellite-assisted MEC network has the best service performance to UEs with the lowest average delay. CSs have powerful storage and data processing capabilities to provide services to all UEs. However, due to the sparse distribution of the CSs on the ground, it does not provide enough coverage to UEs located far away. When only the CSs are relied upon in the MEC network, the service performance of the CSs is affected because all the UEs' task requests are processed by the CSs, which will lead to the overload of the CSs, and the service bandwidth to the UEs becomes smaller. In contrast, the SAT space caching, as the edge nodes, can provide better service to the remote UEs because of the large number of satellites and wide coverage, although their storage and computation capacity are limited. Moreover, since each satellite serves only a certain task type, its load is relatively small, so it can provide a large service bandwidth to the UE. So when there are only SATs in the MEC network, the performance of a large number of SATs jointly providing services to all UEs is better than the case when there are only sparse CSs on the ground. }

\begin{table}[H]
\caption{5 LEO satellite constellations.}
\scalebox{0.85}{
\centering
\begin{tabular}{c|c|cc}
\hline
\textbf{LEO constellation}& \textbf{Altitude (\rj{k}m)} & \multicolumn{2}{c}{\textbf{Number of satellites}} \\ \hline
Starlink & 550                   & \multicolumn{1}{c|}{1584}    & 12000 (\blue{future plan})   \\ \hline
OneWeb            & 1200                  & \multicolumn{1}{c|}{716}     & 6372 (\blue{future plan})             \\ \hline
Amazon            & 630                   & \multicolumn{1}{c|}{578}     & 3236 (\blue{future plan})         \\ \hline
Telesat phase 1   & 1015                  & \multicolumn{1}{c|}{298}     & -                 \\ \hline
Telesat phase 2   & 1325                  & \multicolumn{1}{c|}{1671}    & -   \\ \hline
\end{tabular}
}
\label{5 LEO constellation}
\end{table}

\begin{figure}[]
    \centering   \includegraphics[width=0.4\textwidth]{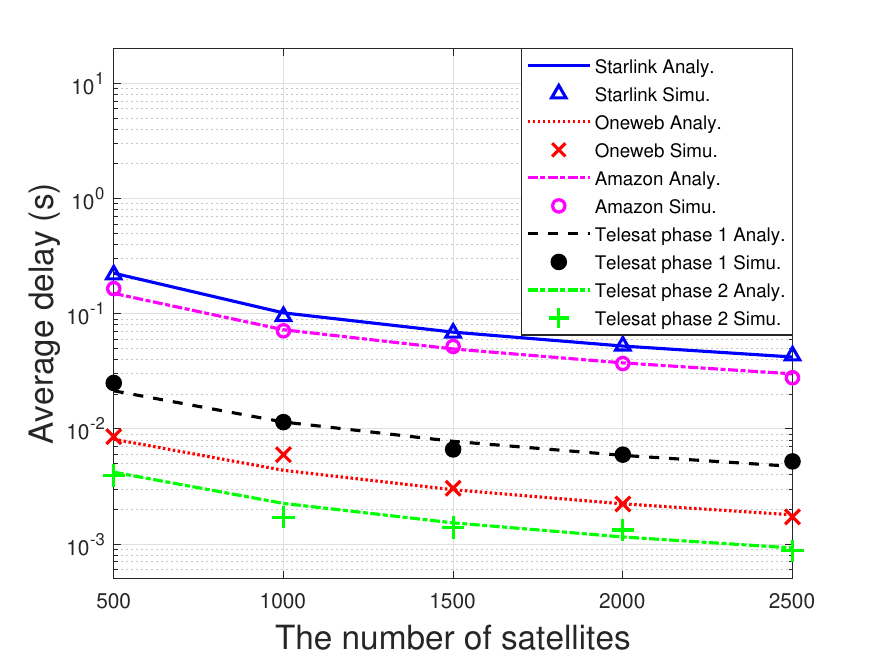}
\caption{\Riotj{Comparison of average delay for different LEO SAT constellations.}}
    \label{Comparison of average delay for different satellite parameters.}
\end{figure}
\Riotj{We extend the analytical framework of this paper to the analysis of the MEC network with well-known LEO satellite constellations.} The parameters of \blue{five commonly known} LEO satellite constellations are given in Table \ref{5 LEO constellation} referring to \cite{9473799}.  \Riotj{Fig. \ref{Comparison of average delay for different satellite parameters.} shows the average delay performance of applying the realistic LEO satellite constellation to the MEC network. }The average delay decreases as the number of SATs increases while other parameters remain constant. \rj{It is possible to enhance the delay performance of MEC network for \iotj{UE}s by increasing the number and height of SATs. }\Riotj{For LEO satellite constellation design, when the LEO satellite orbit altitude is not easy to change, we can increase the number of satellites to reduce the average delay of the MEC network. Meanwhile, for different satellite constellations with the same number of satellites, the delay performance of the MEC network employing a higher orbit LEO satellite constellation performs better.}

\begin{figure*}[ht]
    \centering
% \begin{subfigure}
\begin{minipage}{0.32\textwidth}
    % \subfigure[Effect of $\lambda_u$ with 500 km of SAT]{%[width= \linewidth]
    \centering
    \includegraphics[width=0.98\textwidth]{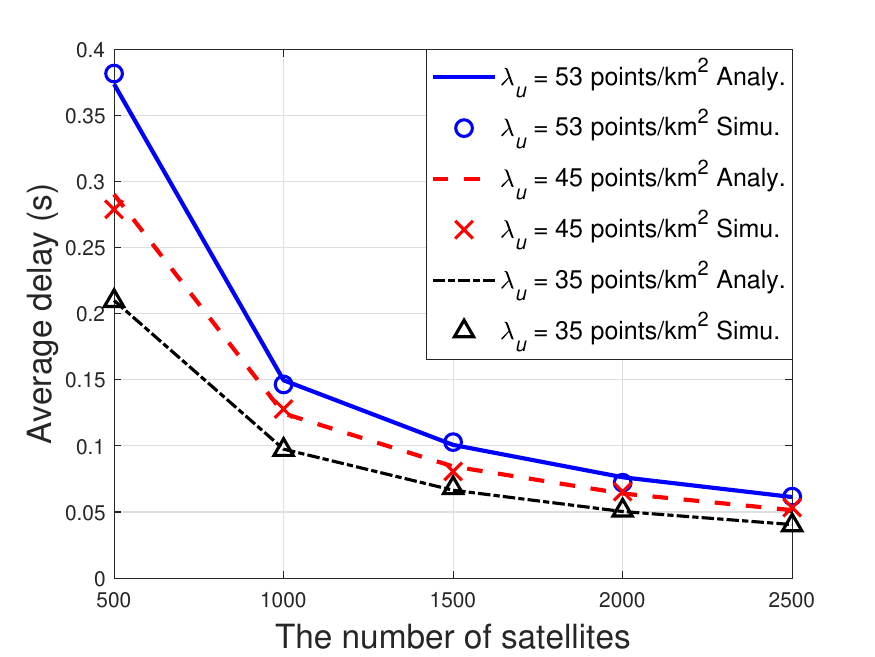}
    \caption{Effect of $\lambda_u$ with 500 km of SAT.}
    \label{Effect of lambda_u with 500 km of SAT}%}
    
\end{minipage}
% \hfill
\begin{minipage}{0.32\textwidth}
     % \subfigure[Effect of $\lambda_u$ with 800 km of SAT]{
    \centering
    \includegraphics[width=0.98\textwidth]{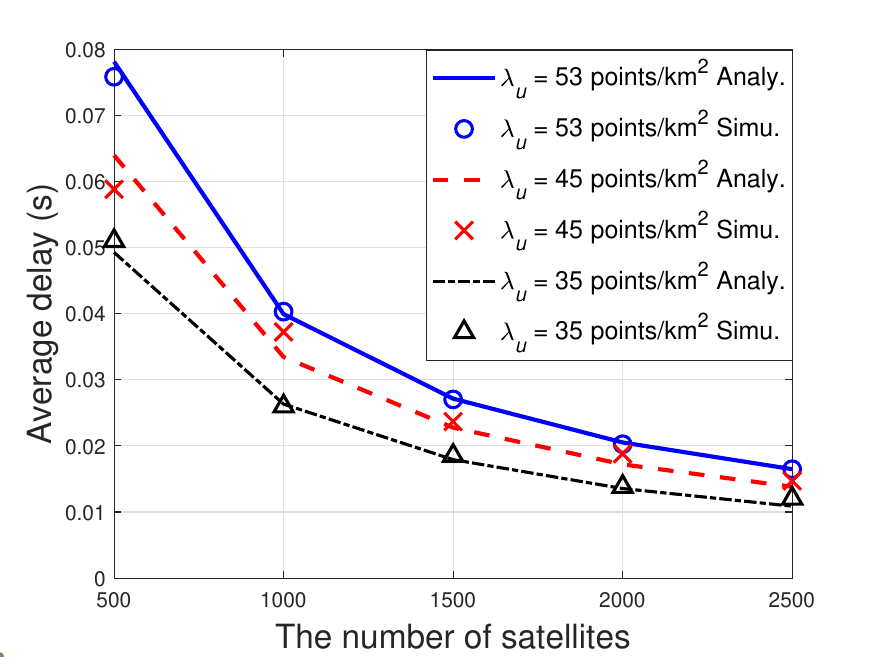}
\caption{Effect of $\lambda_u$ with 800 km of SAT.}
 \label{Effect of lambda_u with 800 km of SAT}%}
 \end{minipage}
 % \hfill
 \begin{minipage}{0.32\textwidth}
   % \subfigure[Effect of $\lambda_u$ with 1000 km of SAT]{%[width= \linewidth]
    \centering
    \includegraphics[width=0.98\textwidth]{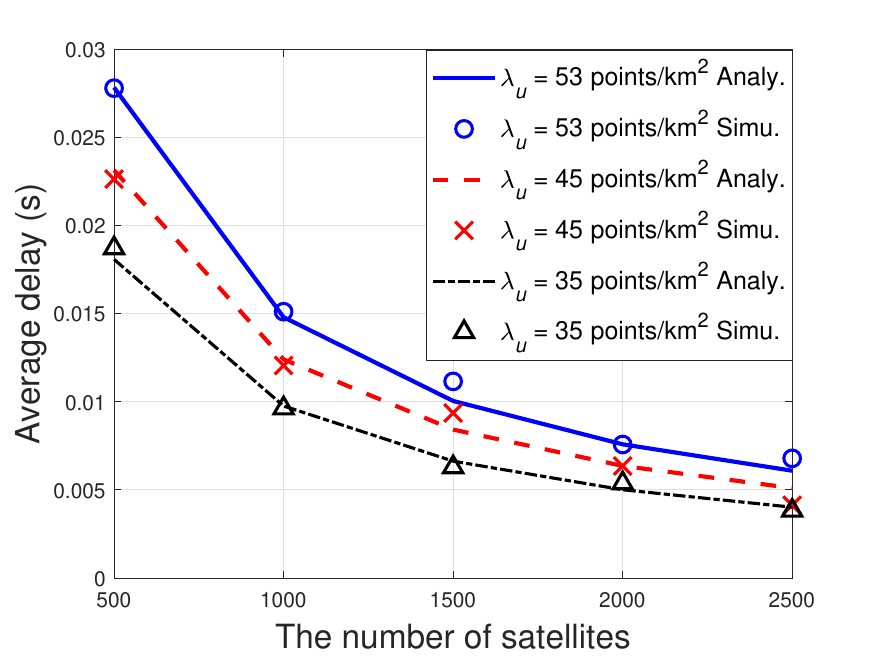}
    \caption{Effect of $\lambda_u$ with 1000 km of SAT.}
    \label{Effect of lambda_u with 1000 km of SAT}%}
    % \caption{}
\end{minipage}
\end{figure*}

\aciotj{Fig.~\ref{Effect of lambda_u with 500 km of SAT}--\ref{Effect of lambda_u with 1000 km of SAT} illustrate how the average delay varies with the number of SATs for different UE densities at three SAT altitudes (500 km, 800 km and 1000 km). As shown in these figures, increasing the number of SATs reduces the average delay across different UE densities, indicating that increasing the number of SATs can enhance the system performance. Furthermore, when the SAT altitude remains constant, it can be observed that the average delay rises with increasing the UE density. }

\begin{figure*}
    \centering
% \begin{subfigure}
\begin{minipage}{0.32\textwidth}
    % \subfigure[Effect of $\lambda_u$ with 500 km of SAT]{%[width= \linewidth]
    \centering
    \includegraphics[width=0.98\textwidth]{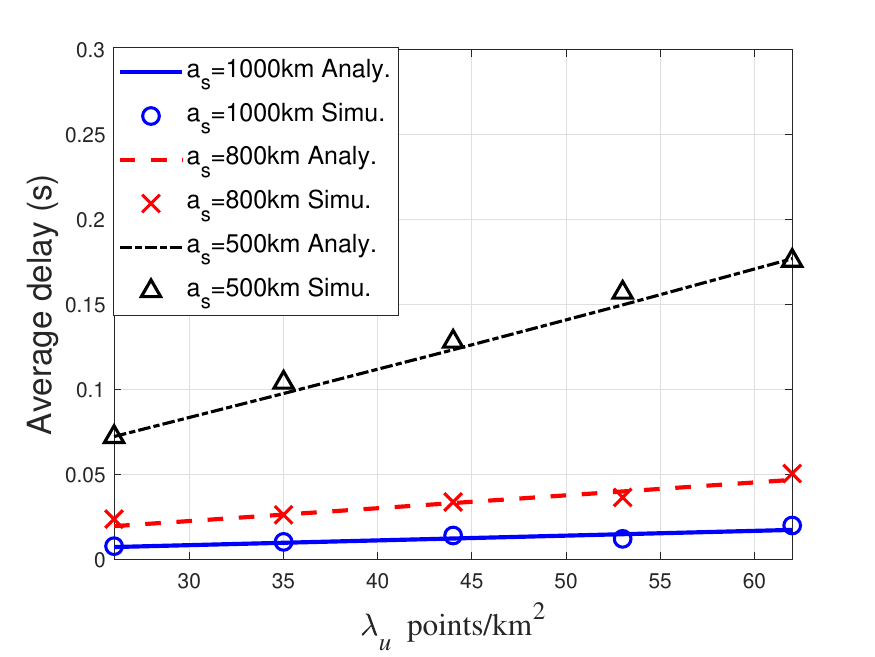}
    \caption{Effect of $\lambda_u$ with 1000 SATs.}
     \label{Effect of lambda_u with 1000 SATs.}
\end{minipage}
% \hfill
\begin{minipage}{0.32\textwidth}
     % \subfigure[Effect of $\lambda_u$ with 800 km of SAT]{
    \centering
    \includegraphics[width=0.98\textwidth]{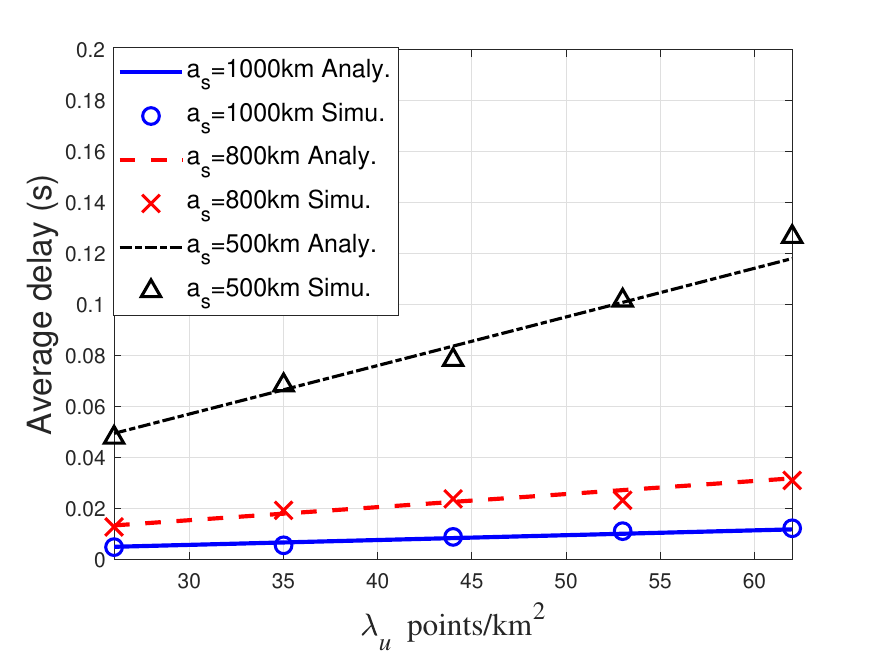}
\caption{Effect of $\lambda_u$ with 1500 SATs.}
 \label{Effect of lambda_u with 2200 SATs.}%}
 \end{minipage}
 % \hfill
 \begin{minipage}{0.32\textwidth}
   % \subfigure[Effect of $\lambda_u$ with 1000 km of SAT]{%[width= \linewidth]
    \centering
    \includegraphics[width=0.98\textwidth]{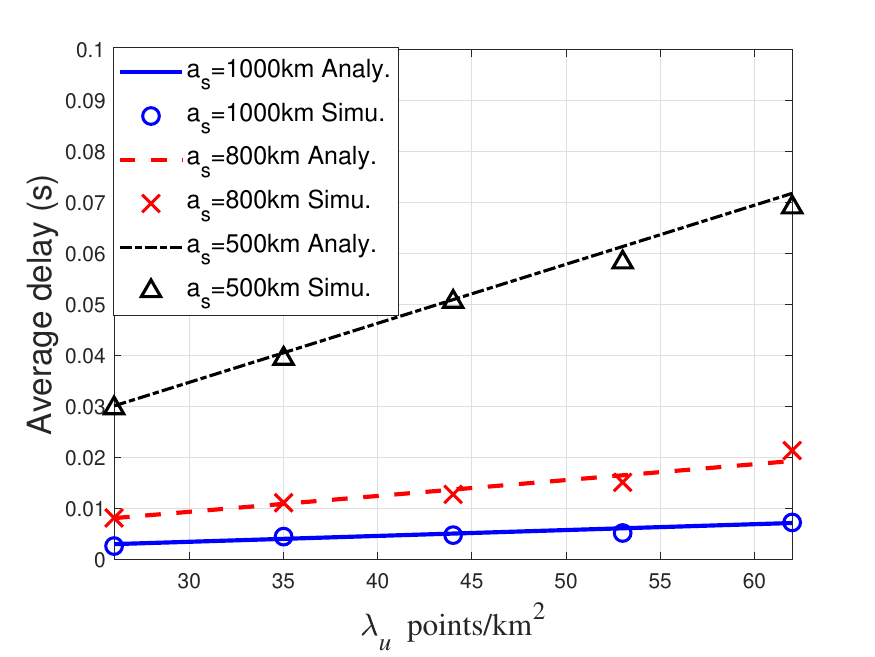}
    \caption{Effect of $\lambda_u$ with 2500 SATs.}
    \label{Effect of lambda_u with 3600 SATs.}%}
    % \caption{}
\end{minipage}
\end{figure*}

\aciotj{Next, Fig.~\ref{Effect of lambda_u with 1000 SATs.}--\ref{Effect of lambda_u with 3600 SATs.} further illustrate the relationship between the average delay and UE density for different numbers and altitudes of SATs. Each figure corresponds to a specific number of SATs (1000, 1500, and 2500) and shows the average delay versus the UE density for three different SAT altitudes. The three curves in each figure represent the average delay at three altitudes. As the UE density increases, the average delay at all three altitudes exhibits an increasing trend, consistent with Fig.~\ref{Effect of lambda_u with 500 km of SAT}--\ref{Effect of lambda_u with 1000 km of SAT}. Further analysis of Fig.~\ref{Effect of lambda_u with 1000 SATs.}--\ref{Effect of lambda_u with 3600 SATs.} reveals that at the SAT altitude of 500 km, the average delay changes more significantly as the UE density rises, and the growth of the average delay gradually slows down with increasing SAT altitude.} It can be seen that the higher the SAT altitude, the greater the number of \iotj{UE}s that can be served within a specific range. \rj{In order to minimise costs while ensuring network performance, it is not necessary to deploy an excessive number of SATs in areas with sparse \iotj{UE}s. In contrast, in areas with higher \iotj{UE} density, more SATs typically reduce the average delay. However, in order to save costs, it is possible to opt for fewer SATs with increased altitude. This way can also maintain or optimise the quality of network service even with a small number of SATs.}
 
% \vspace{-4mm}
\section{Conclusion}
% \vspace{-1mm}
In this paper, we propose a MEC network with space caching to enhance the performance of the network system. The SAT and CS are utilized to provide \rj{service for the \iotj{UE} task}, and the association strategy and service model between \iotj{UE} with SAT and CS are modeled. The analytical expression for the uplink and downlink coverage probabilities are derived \rj{using stochastic geometry tools}, and we also calculate the average delay of the system \rj{based on queuing theory, considering the different task and service types}. \taes{This work is the first attempt to analyse the performance of the LEO satellite-assisted space caching network using stochastic geometry, and the analytical expressions obtained are tractable for performance analysis. }\taes{The system performance is simulated and analysed, and the correctness of the system performance analysis is verified through Monte Carlo simulation.} We analyze the impact of the parameters on the system performance \rj{in terms of delay} and demonstrate the \blue{performance gains from increasing the number of SATs or increasing their altitude on reducing the average delay.} \taes{For regions with different \iotj{UE} densities, the optimal approach to maintain network performance while still saving SAT deployment costs is to appropriately increase the SAT altitude, which could reduce the need for the number of SATs to a greater extent. The average latency of the LEO satellite-assisted space caching MEC network outperforms that of the MEC network with CS only and SAT only, which could} \rj{prove the potential of SAT \Riotj{seamless coverage characteristic} to improve system performance in the MEC network}. \taes{The \mktaes{analytical} results provide useful instructions for the configuration and planning of SAT constellations for MEC networks with space caching.} \taes{Based on the above analysis, these system-level insights into the network design could provide useful guidance for further improving the quality of service to \iotj{UE}s.} 

\Riotj{This work can be extended in several directions, such as interference and energy efficiency studies. In our future work, we will further investigate the interference and energy efficiency in LEO satellite-assisted MEC networks based on this work. }

% \vspace{-3mm}
\appendices
\section{Proof of Lemma \ref{lemma SU CP}}
% \vspace{-2mm}
According to the definition of (\ref{defin CP SU}), the downlink coverage probability of the S-U link is the probability that the S-U link \blue{${\mathrm{SNR}_{S_i}^D}$} is greater than the threshold $\tau$ within the $i$ tier.
% \footnotesize\begin{align}
% \begin{small}
% \begin{equation}
\begin{align}
% \begin{aligned}
&\quad\blue{P_{{S_i}}^D( \tau  ) = \mathbb{P} \big[ {\mathrm{SNR}_{{S_i}}^D > \tau } \big] }\nonumber\\
& = \mathbb{P}\Big[ {\frac{{p_s( {\frac{c}{{4\pi {f_s}}}} )^2{{\left| {{h_0}} \right|}^2}{{ \blue{D_{d,S-U}^{-2}}}}}}{{\sigma _U^2}} > \tau }\Big] \nonumber\\
& = {\mathbb{E}_{\blue{D_{d,S-U}}}}\Big[ \mathbb{P} {({{{\left| {{h_0}} \right|}^2} > \frac{{\tau \sigma _U^2\blue{D_{d,S-U}^{2}}}}{{p_s( {\frac{c}{{4\pi {f_s}}}} )^2}}\left| {\blue{D_{d,S-U}} = x} \right.})} \Big]\nonumber \\
& \mathop = \limits^{\left( a \right)} {\mathbb{E}_{\blue{D_{d,S-U}}}}\left[ \mathbb{P}{( {{{\left| {{h_0}} \right|}^2} > A\times {\blue{D_{d,S-U}^{2}}}\left| {\blue{D_{d,S-U}} = {x}} \right.} )} \right] \nonumber\\
& = {\mathbb{E}_{\blue{D_{d,S-U}}}}\left[ {1 - {F_{{{\left| {{h_0}} \right|}^2}}}( {A\times{\blue{D_{d,S-U}^{2}}}} )\left| {\blue{D_{d,S-U}} = x} \right.} \right] \nonumber\\
& = \int_{{a_s}}^{{d_{\max }}} {( {1 - {F_{{{\left| {{h_0}} \right|}^2}}}( {Ax^2} )} )} {f_{\blue{D_{d,S-U}}}}\left( {x} \right)\mathrm{d}x\nonumber \\
& \mathop  = \limits^{\left( b \right)} \int_{{a_s}}^{{d_{\max }}} {\frac{{\blue{N_{i}}}}{{{\pi}{r_e}{r_s}}}( {1 - {F_{{{\left| {{h_0}} \right|}^2}}}( {Ax^2})} )} \frac{{x}}{{\sqrt {1 \!-\! {{( {1 \!-\!\frac{{x^2 - a_s^2}}{{2{r_e}{r_s}}}} )}^2}} }} \nonumber\\
& \quad \!\times\! {\Big[ {1 \!-\! \frac{1}{{{\pi}}}\arccos ( {1 \!-\!\frac{{x^2 - a_s^2}}{{2{r_e}{r_s}}}} )} \Big]^{\blue{N_{i}} \!-\! 1}}\mathrm{d}x\nonumber \\
& \mathop  = \limits^{\left( c \right)} \frac{\blue{N_{i}}}{{{\pi}{r_e}{r_s}}}\int_{{a_s}}^{{d_{\max }}} {\Big[ {1\!-\! {(1 \!-\! {e^{ - \frac{{\mu A{x^2}}}{\beta _s}}})^{\alpha _s}}} \Big]} \frac{{x}}{{\sqrt {1 \!-\!{{( {1\!-\! \frac{{x^2 - a_s^2}}{{2{r_e}{r_s}}}})}^2}} }} \nonumber\\
& \quad\!\times\! {\Big[ {1 \!-\! \frac{1}{{{\pi}}}\arccos ( {1 \!-\!\frac{{x^2 - a_s^2}}{{2{r_e}{r_s}}}} )} \Big]^{\blue{N_{i}} - 1}}\mathrm{d}x \nonumber\\
& \mathop  = \limits^{\left( d \right)} \frac{\blue{N_{i}}}{{{\pi }{r_e}{r_s}}}\int_{{a_s}}^{{d_{\max }}} {\Big[ {1 \!-\! ( {\sum\limits_{j = 0}^{{\alpha _s}} {( {_j^{{\alpha _s}}} )} {{\left( { - 1} \right)}^j}\exp ( - \frac{{j\mu Ax^2}}{{{\beta _s}}})} )} \Big] } \nonumber\\
& \quad\!\times\! \frac{{x}}{{\sqrt {1\!-\!{{( {1 \!-\! \frac{{x^2 - a_s^2}}{{2{r_e}{r_s}}}} )}^2}} }}{{\Big[ {1\!-\! \frac{1}{{{\pi}}}\arccos( {1 \!-\! \frac{{x^2 - a_s^2}}{{2{r_e}{r_s}}}} )} \Big]}^{{\blue{N_{i}}}\!-\! 1}} \mathrm{d}x\nonumber \\
& \mathop  =  \frac{{\blue{N_{i}}}}{{{\pi}{r_e}{r_s}}}\int_{{a_s}}^{{d_{\max }}} \Big[ {\sum\limits_{j = 1}^{{\alpha _s}} {\left( {_j^{{\alpha _s}}} \right)} {{( { - 1} )}^{j + 1}}\exp ( - \frac{{j\mu Ax^2}}{{{\beta _s}}})} \Big] \nonumber\\
& \quad\!\times\! \frac{x}{{\sqrt {1 \!-\! {{( {1 \!-\!\frac{{x^2 - a_s^2}}{{2{r_e}{r_s}}}} )}^2}} }}{{\Big[ {1 - \frac{1}{{{\pi}}}\arccos ( {1\!-\! \frac{{x^2 \!-\! a_s^2}}{{2{r_e}{r_s}}}} )} \Big]}^{{\blue{N_{i}}} \!-\! 1}} \mathrm{d}x,
\end{align}
where ${d_{\max }}= \sqrt {2r_ea_s + a_s^2}$, $\mu  = {\left( {{\alpha _s!}} \right)^{ - \frac{1}{{{\alpha _s}}}}}$, $\sigma _U^2$ is the noise variance of the \iotj{UE}. \rj{(a) is obtained by letting $A = \frac{{\tau \sigma _U^2}}{{\blue{p_s}( {\frac{c}{{4\pi {f_s}}}} )^2}}$. (b) is derived from (\ref{S-U pdf}). (c) follows from the tight bound assumption of CDF of Gamma distribution, as stated in \eqref{app 2}. (d) follows from the binomial theorem with the assumption that ${\alpha _s} \in \mathbb{N}$. }

% \vspace{-4mm}
 \section{Proof of Lemma \ref{lemma AP}}
% \vspace{-2mm}
Assuming that $D_k, k \in \left\{ {c,s} \right\}$ is the closest distance from the \iotj{UE} to the nearest server in the $k$ tier. The nearest server SAT or CS is chosen to provide the maximum average power for connecting. $B_k$ represents the bias factor of the $k$-tier servers.

According to the association probability equation ${\cal A}_k$ in (\ref{defin AP}), we can make the following calculations 
\begin{align}
% \setlength\abovedisplayskip{3pt}
% \setlength\belowdisplayskip{3pt}
% \begin{array}
& \quad\mathbb{P}\big[{p_s}{B_s}{( {\frac{c}{{4\pi {f_s}{D_s}}}} )^2} \geq {p_{c}}{B_{c}}\bk{(\frac{c}{4\pi f_c D_c})}^{ \alpha }\big]\nonumber \\
& = \mathbb{P}\big[D_c\geq {( {\frac{{{p_c}{B_c}}}{{{p_s}{B_s}}}} )^{\frac{1}{\alpha }}}{( {\frac{{4\pi {f_s}}}{c}} )^{\frac{2}{\alpha }}}\bk{(\frac{c}{4\pi f_c})}D_s^{\frac{2}{\alpha }}\left| {_{{D_s} = x}} \big] \right.,
% \end{array}
\end{align}
and
\begin{align}
% \setlength\abovedisplayskip{3pt}
% \setlength\belowdisplayskip{3pt}
    % \begin{array}{l}
&\quad \mathbb{P}\big[{p_c}{B_c}\bk{(\frac{c}{4\pi f_c D_c})}^{ \alpha } > {p_s}{B_s}{( {\frac{c}{{4\pi {f_s}{D_s}}}})^2}\big] \nonumber\\
& = \mathbb{P} \big[ D_s > {( {\frac{{{p_s}{B_s}}}{{{p_c}{B_c}}}} )^{\frac{1}{2}}}( {\frac{c}{{4\pi {f_s}}}} )\bk{(\frac{c}{4\pi f_c})^{-\frac{\alpha}{2}}}D_c^{\frac{\alpha }{2}}\left| {_{{D_c} = x}} \big] \right..
% \end{array}
\end{align}

The association probability expression of the $k$ tier is 
\begin{align}
    % \begin{array}{l}
{{\cal A}_k} = \begin{cases}
\mathbb{P}[ {D_{c} \geq {{( {\frac{{{p_{c}}{B_{c}}}}{{{p_s}{B_s}}}} )}^{\frac{1}{\alpha }}}{{( {\frac{{4\pi {f_s}}}{c}} )}^{\frac{2}{\alpha }}}{\bk{(\frac{c}{4\pi f_c })}}x^{\frac{2}{\alpha }} }], &  k = {s} \\
\mathbb{P}[ {D_s > {{( {\frac{{{p_s}{B_s}}}{{{p_{c}}{B_{c}}}}} )}^{\frac{1}{2}}}( {\frac{c}{{4\pi {f_s}}}} ){\bk{(\frac{c}{4\pi f_c })}^{-\frac{\alpha}{2}}}x^{\frac{\alpha }{2}}} ],& k = c
 \end{cases}.
% \end{array}
\end{align}

The association probabilities of the SAT and CS are solved separately below.

Considering the different task types in the network, we assume a total of ${\cal M}+1$ tiers in this network, where CS is tier $0$ and SAT of type-$i$ offloadable is tier $i$ ($1\leq i\leq {\cal M}$). ${{\cal A}_{\blue{i}}}$ means that the \iotj{UE} associates with the nearest type-$i$ offloadable SAT 
\begin{align}
{{\cal A}_{{i}}}& = \mathbb{P} \big[ {{p_s}{B_s}{{( {\frac{c}{4\pi {f_s}{D_\blue{s_i}}}} )}^2} \geq {p_c}{B_c}\bk{(\frac{c}{4\pi f_c D_c})}^{ \alpha }}\big]\nonumber\\
& = \mathbb{P} \big[ {D_c\geq {{( {\frac{{{p_c}{B_c}}}{{{p_s}{B_s}}}} )}^{\frac{1}{\alpha }}}{{( {\frac{{4\pi {f_s}}}{c}} )}^{\frac{2}{\alpha }}}{\bk{(\frac{c}{4\pi f_c })}}D_\blue{s_i}^{\frac{2}{\alpha }}} \left| {_\blue{{D_{s_i}} = x}}   \right.\big]\nonumber\\
& \mathop = \limits^{( a)} \mathbb{P} \big[ {D_c\geq {Q_s}D_\blue{s_i}^{\frac{2}{\alpha }}} \left| {_\blue{{D_{s_i}} = x}}\right.  \big]\nonumber\\
&= \int_0^\infty {{\mathbb{P}}[ {D_c\geq {Q_s}x^{\frac{2}{\alpha }}} ]} {f_{{D_\blue{s_i}}}}( x )\mathrm{d}x\nonumber\\
& = \int_0^\infty  {(1 - {F_{{D_{c}}}}({Q_s}x^{\frac{2}{\alpha }}))} {f_{{D_\blue{s_i}}}}( x )\mathrm{d}x,
\end{align}
where, (a) is letting ${Q_{s}} = {( {\frac{{{p_c}{B_c}}}{{{p_{{s}}}{B_{{s}}}}}} )^{\frac{1}{\alpha }}}{( {\frac{{4\pi {f_s}}}{c}} )^{\frac{2}{\alpha }}}{\bk{(\frac{c}{4\pi f_c })}}$.

According to (\ref{S-U pdf}), it is known ${f_{{D_\blue{s_i}}}}\left( x \right)$ that
% \begin{small}
\begin{align}
&\quad{f_{{D_\blue{s_i}}}}\left( x \right)  \nonumber\\
&= \begin{cases}
\frac{{{N_\blue{i}}}}{{\pi {r_e}{r_s}}} \times \frac{x}{{\sqrt {1 - {( {1 - \frac{{{x^2} - a_s^2}}{{2{r_e}{r_s}}}} )^2}} }}\\
\quad{\times[ {1 \!-\! \frac{1}{\pi }\arccos( {1 \!-\! \frac{{{x^2} - a_s^2}}{{2{r_e}{r_s}}}} )} ]^{{N_\blue{i}} \!-\! 1}},& {a_s} \le x < {d_{\max }}\\
0,& else
\end{cases},
\end{align}
% \end{small}
where, $r_e$ is the radius of the earth, $a_s$ is the height of the SAT from the \rj{E}arth.

According to (\ref{CDF C-U}), we can get the ${\overline F _{{D_{c}}}}({Q_s}x^{\frac{2}{\alpha }})$ is ${\overline F _{{D_{c}}}}({Q_s}x_{}^{\frac{2}{\alpha }}) = \exp ( { - {\lambda _c}\pi Q_s^2x_{}^{\frac{4}{\alpha }}} )$, ${d_{\max }} = \sqrt {2{r_e}{a_s} + a_s^2} $.

\begin{itemize}
\item  When $x<a_s$
\begin{align}
{{\cal A}_\blue{i}}  = \int_0^\infty  {(1 - {F_{{D_{c}}}}({Q_s}x_{}^{\frac{2}{\alpha }}))} {f_{{D_\blue{s_i}}}}\left( x \right)dx = 0.
\label{AS-1}
\end{align}

\item When ${a_s} \le x < {d_{\max }}$
\begin{align}
&\quad{{\cal A}_{_\blue{i}}} = \int_0^\infty  {(1 - {F_{{D_{c}}}}({Q_s}x_{}^{\frac{2}{\alpha }}))} {f_{{D_\blue{s_i}}}}\left( x \right)\mathrm{d}x\nonumber\\
& = \frac{{{N_\blue{i}}}}{{\pi {r_e}{r_s}}}\int_{{a_s}}^{{d_{\max }}} {\exp ( { - {\lambda _c}\pi Q_s^2x_{}^{\frac{4}{\alpha }}} )}  \frac{x}{{\sqrt {1 - {{( {1 - \frac{{{x^2} - a_s^2}}{{2{r_e}{r_s}}}} )}^2}} }}\nonumber\\
&\quad\!\times \!{\big[ {1 - \frac{1}{\pi }\arccos ( {1 - \frac{{{x^2} - a_s^2}}{{2{r_e}{r_s}}}})} \big]^{{N_\blue{i}} - 1}}\mathrm{d}x.
\label{AS-2}
\end{align}

\item When ${d_{\max }} \le x$
\begin{align}
{{\cal A}_{_\blue{i}}} = \int_0^\infty  {(1 - {F_{{D_{c}}}}({Q_s}x_{}^{\frac{2}{\alpha }}))} {f_{{D_\blue{s_i}}}}( x )\mathrm{d}x= 0.
\label{AS-3}
\end{align}
 \end{itemize}
Considering (\ref{AS-1}), (\ref{AS-2}) and (\ref{AS-3}), we could get the association probability for SAT as follows
\begin{align}
{{\cal A}_\blue{i}} = \frac{{{N_\blue{i}}}}{{\pi {r_e}{r_s}}}\int_{{a_s}}^{{d_{\max }}} {\exp ( { - {\lambda _c}\pi Q_s^2x^{\frac{4}{\alpha }}} )} \frac{x}{\sqrt {1 - {( {1 - \frac{{{x^2} - a_s^2}}{{2{r_e}{r_s}}}})^2}}}\nonumber\\
{\times[{1 - \frac{1}{\pi }\arccos ( {1 - \frac{{{x^2} - a_s^2}}{{2{r_e}{r_s}}}})}]^{{N_\blue{i}} - 1}}\mathrm{d}x,
\label{AS}
\end{align}
where ${Q_s} = {( {\frac{{{p_c}{B_c}}}{{{p_s}{B_s}}}} )^{\frac{1}{\alpha }}}{( {\frac{{4\pi {f_s}}}{c}})^{\frac{2}{\alpha }}}{\bk{(\frac{c}{4\pi f_c })}}$, ${d_{\max }} = \sqrt {2{r_e}{a_s} + a_s^2} $.

${\cal A}_\blue{0}$ means that the \iotj{UE} associates with the nearest CS  
\begin{align}
{{\cal A}_\blue{0}}& = \mathbb{P}\big[ {{p_c}{B_c}\bk{(\frac{c}{4\pi f_c D_c})}^{ \alpha }> {p_s}{B_s}{{( {\frac{c}{{4\pi {f_s}{D_\blue{s_i}}}}})^2}}}\big]\nonumber\\
& = \mathbb{P}\big[ {D_\blue{s_i} > {( {\frac{{{p_s}{B_s}}}{{{p_c}{B_c}}}} )^{\frac{1}{2}}}( {\frac{c}{{4\pi {f_s}}}})\bk{(\frac{c}{4\pi f_c})^{-\frac{\alpha}{2}}}D_c^{\frac{\alpha }{2}}}\blue{\left| {_{{D_{c}} = x}}   \right.}\big]\nonumber\\
& \mathop = \limits^{( a)}\mathbb{P}\big[ {D_\blue{s_i} > Q_s^{ - \frac{\alpha }{2}}D_c^{\frac{\alpha }{2}}}\blue{\left| {_{{D_{c}} = x}}   \right.} \big]\nonumber\\
& = \int_0^\infty {\mathbb{P}[{D_\blue{s_i}> Q_s^{ - \frac{\alpha }{2}}x^{\frac{\alpha }{2}}} ]} {f_{{D_{c}}}}( x )\mathrm{d}x \nonumber\\
&  = \int_0^\infty  {(1 - {F_{D_\blue{s_i}}}(Q_s^{ - \frac{\alpha }{2}}x_{}^{\frac{\alpha }{2}})} {f_{{D_{c}}}}\left( x \right)\mathrm{d}x,
\end{align}
where (a) is derived by letting $Q_s^{ - \frac{\alpha }{2}} = {( {\frac{{{p_s}{B_s}}}{{{p_c}{B_c}}}} )^{\frac{1}{2}}}( {\frac{c}{{4\pi {f_s}}}} )\bk{(\frac{c}{4\pi f_c})^{-\frac{\alpha}{2}}}$.

As for CS, we could get ${f_{{D_{c}}}}$ is ${f_{{D_{c}}}}( x ) = 2{\pi} {\lambda _c}x{\exp ( { - \pi {\lambda _c}{x^2}})}$ by (\ref{PDF C-Ueq}).

According to (\ref{CDF S-U}), we could get the CDF of distance distribution ${F_{{D_\blue{s_i}}}}(Q_s^{ - \frac{\alpha }{2}}{x^{\frac{\alpha }{2}}})$ as follow
\begin{align}
&F_{{D_\blue{s_i}}}(Q_s^{ - \frac{\alpha}{2}}{x^{\frac{\alpha }{2}}}) = \nonumber\\
&\begin{cases}
0, & \hspace{-19mm} x< {( {{a_s}Q_s^{\frac{\alpha}{2}}} )^{\frac{2}{\alpha }}}\\
1 \!-\! \big[ {1 \!-\! {\frac{1}{\pi}}\arccos(1\! - \!\frac{{(Q_s^{\!- {\frac{\alpha }{2}}}{x^{\frac{\alpha }{2}}})^2} - a_s^2}{2{r_e}{r_s}} )} \big]^{N_\blue{i}}\!,\\
& \hspace{-19mm}( {{a_s}Q_s^{\frac{\alpha }{2}}})^{\frac{2}{\alpha }}\!\! \leq x <\! {( {{d_{\max }}Q_s^{\frac{\alpha }{2}}} )^{\frac{2}{\alpha }}} \\
1\! -\! {\big[ {1\! -\! \frac{1}{\pi }\arccos ({\frac{{r_e}}{r_s}} )} \big]^{N_\blue{i}}}, &\hspace{-19mm}  {( {{d_{\max}}Q_s^{\frac{\alpha }{2}}} )^{\frac{2}{\alpha }}} \leq x    
\end{cases},
\end{align}
where ${d_{\rm max }} = \sqrt {2{r_e}{a_s} + a_s^2} $.

\begin{itemize}

\item When $x < {( {{a_s}Q_s^{\frac{\alpha }{2}}} )^{\frac{2}{\alpha }}}$
\begin{align}
&{{\cal A}_\blue{0}}  = \int_0^\infty  {(1 - {F_{D_\blue{s_i}}}(Q_s^{ - \frac{\alpha }{2}}x_{}^{\frac{\alpha }{2}})} {f_{{D_{c}}}}(x)\mathrm{d}x \nonumber\\
& = \int_0^{{{({a_s}Q_s^{\frac{\alpha }{2}})}^{\frac{2}{\alpha }}}} {2\pi {\lambda _c}x \cdot \exp( { - \pi {\lambda _c}{x^2}} )\mathrm{d}x}.
\label{AC-1}
\end{align}

\item When ${( {{a_s}Q_s^{\frac{\alpha }{2}}} )^{\frac{2}{\alpha }}}{\kern 1pt}  \le x < {( {{d_{\max }}Q_s^{\frac{\alpha }{2}}} )^{\frac{2}{\alpha }}}$
\begin{small}
\begin{align}
&{{\cal A}_\blue{0}} = \int_0^\infty  {(1 - {F_{{D_\blue{s_i}}}}(Q_s^{ - \frac{\alpha }{2}}x^{\frac{\alpha }{2}})} {f_{{D_{c}}}}(x)\mathrm{d}x\nonumber \\
& = \int_{{{({a_s}Q_s^{\frac{\alpha }{2}})}^{\frac{2}{\alpha }}}}^{{{({d_{\max }}Q_s^{\frac{\alpha }{2}})}^{\frac{2}{\alpha }}}} {{\big[ {1 - \frac{1}{\pi }\arccos ( {1 - \frac{{{{(Q_s^{ - \frac{\alpha }{2}}x_{}^{\frac{\alpha }{2}})}^2} - a_s^2}}{{2{r_e}{r_s}}}} )} \big]^{N_\blue{i}}}}\nonumber\\
&\quad\times2\pi {\lambda _c}x \exp ( { - \pi {\lambda _c}{x^2}} )\mathrm{d}x\nonumber\\
& =  2\pi {\lambda _c}\int_{{{({a_s}Q_s^{\frac{\alpha }{2}})}^{\frac{2}{\alpha }}}}^{{{({d_{\max }}Q_s^{\frac{\alpha }{2}})}^{\frac{2}{\alpha }}}}\!\!\! {\big[ {1 - \frac{1}{\pi }\arccos ( {1 - \frac{{{{(Q_s^{ - \frac{\alpha }{2}}x_{}^{\frac{\alpha }{2}})}^2} - a_s^2}}{{2{r_e}{r_s}}}} )} \big]^{N_\blue{i}}} \nonumber\\
& \quad\times x  \exp ( { - \pi {\lambda _c}{x^2}} )\mathrm{d}x.
\label{AC-2}
\end{align}
\end{small}

\item When ${( {{d_{\max }}Q_s^{\frac{\alpha }{2}}} )^{\frac{2}{\alpha }}} \le x$
\begin{align}
&{{\cal A}_\blue{0}}  = \int_0^\infty  {(1 - {F_{{D_s}}}(Q_s^{ - \frac{\alpha }{2}}x^{\frac{\alpha}{2}})} {f_{{D_{c}}}}( x )\mathrm{d}x\nonumber \\
& = 2\pi {\lambda _c}\int_{{\kern 1pt} {{({d_{\max }}Q_s^{\frac{\alpha }{2}})}^{\frac{2}{\alpha }}}}^\infty  {{{\big[ {1 - \frac{1}{\pi }\arccos ( {\frac{{{r_e}}}{{{r_s}}}})} \big]^{{N_\blue{i}}}}} x}\nonumber\\
& \quad\times\exp ( { - \pi {\lambda _c}{x^2}} )\mathrm{d}x.
\label{AC-3}
\end{align}
% \end{footnotesize}

\end{itemize}

Considering (\ref{AC-1}), (\ref{AC-2}) and (\ref{AC-3}), the final expression for $\blue{{\cal A}_{0}}$ can be derived
% \begin{footnotesize}
\begin{align}
&{{\cal A}_\blue{0}}= \int_0^\infty  {(1 - {F_{{D_s}}}(Qx))} {f_{{D_{c}}}}\left( x \right)\mathrm{d}x\nonumber \\
& = \int_0^{{{({a_s}Q_s^{\frac{\alpha }{2}})}^{\frac{2}{\alpha }}}} {2\pi {\lambda _c}x \cdot \exp ( { - \pi {\lambda _c}{x^2}} )\mathrm{d}x}  \nonumber\\
&  + 2\pi {\lambda _c}\int_{{{({a_s}Q_s^{\frac{\alpha }{2}})}^{\frac{2}{\alpha }}}}^{{{({d_{\max }}Q_s^{\frac{\alpha }{2}})}^{\frac{2}{\alpha }}}} {\!{{\big[ {1 \!-\! \frac{1}{\pi }\arccos ( {1 \!-\! \frac{{{(Q_s^{ - \frac{\alpha }{2}}x_{}^{\frac{\alpha }{2}})^2} \!-\! a_s^2}}{{2{r_e}{r_s}}}} )} \big]^{N_\blue{i}}}} }\nonumber\\
&\quad \times x\exp( { - \pi {\lambda _c}{x^2}} )\mathrm{d}x \nonumber\\
& +2\pi {\lambda _c}\!\int_{{{({d_{\max }}Q_s^{\frac{\alpha }{2}})}^{\frac{2}{\alpha }}}}^\infty{{{\big[ {1 \!-\! \frac{1}{\pi }\arccos ( {\frac{{{r_e}}}{{{r_s}}}})}\big]^{N_\blue{i}}}} x\exp ( { - \pi {\lambda _c}{x^2}} )\mathrm{d}x },
\label{AC}
\end{align}
% \end{footnotesize}
where $Q_s^{ - \frac{\alpha }{2}} = {( {\frac{{{p_s}{B_s}}}{{{p_c}{B_c}}}} )^{\frac{1}{2}}}( {\frac{c}{{4\pi {f_s}}}})\bk{(\frac{c}{4\pi f_c})^{-\frac{\alpha}{2}}}$, ${d_{\max }} = \sqrt {2{r_e}{a_s} + a_s^2} $.

Combining (\ref{AS}) and (\ref{AC}), this proof is completed.

% \vspace{-2mm}
\section{Proof of Lemma \ref{lemma Pu-s CP}}
% \vspace{-2mm}
The definition of U-S uplink coverage probability is given by
\begin{align}
P_{{S_i}}^U\left( \tau  \right) = w\mathbb{P}( {\mathrm{SNR}_{{S_i}}^U > \tau } ),
\label{U-S CP definition}
\end{align}
where the probability $w$ is derived from the null probability of BPP. The probability $w$ indicates the presence of at least one SAT in the spherical cap ${{\cal{A}}_{cap,{\theta _c}}}$ as shown in Fig. \ref{USsystemmode}. The derivation is as follows  %\usepackage{mathrsfs}
\begin{align}
w &= 1 \!-\! \prod\limits_{{\Phi _{s_i}}} \!{( {1 \!-\! \frac{{{\cal S}( {{{\cal{A}}_{cap,{\theta _c}}}} )}}{{{\cal S}( {{{\cal{A}}_{Earth}}} )}}} )} = 1\! -\! \prod\limits_{{\Phi _{s_i}}} {( {1\! -\! \frac{{2\pi r_e^2( {1 \!-\! \cos {\theta _c}} )}}{{4\pi r_e^2}}} )}\nonumber \\
&= 1 - {( {\frac{{1 + \cos {\theta _c}}}{2}} )^{{N_{\blue{i}}} - 1}}.
\end{align}

Solving for (\ref{U-S CP definition}) as follows
% \begin{equation}
% \scalebox{0.9}{$
\begin{footnotesize}
\begin{align}
&\quad P_{{S_i}}^U\left( \tau  \right) = w \mathbb{P} ( {\mathrm{SNR}_{{S_i}}^U > \tau } )\buildrel \Delta \over = w\mathbb{P} \Big[ {\frac{p_u( {\frac{c}{{4\pi {f_s}}}} )^2{{\left| {{h_0}} \right|}^2}D^{-2}_\blue{u,U-S}}{\sigma _S^2} > \tau} \Big]\nonumber\\
 &= w\mathbb{P} \Big[ {{{\left| {{h_0}} \right|}^2} > \frac{{\tau \sigma _S^2{D^{2}_\blue{u,U-S}}}}{{\blue{p_u}( {\frac{c}{{4\pi {f_s}}}} )^2}}}\Big]\nonumber\\
&\mathop  = \limits^{( a)} w{\mathbb{E} _{D_u}}\Big[ \mathbb{P} {( {{{\left| {{h_0}} \right|}^2} > A{D^{2}_\blue{u,U-S}}\left| {D_\blue{u,U-S} = {x}} \right.})} \Big]\nonumber\\
&\mathop  = \limits^{( b )} w{\mathbb{E}_{D_u}}\Big[{1 - {{\big( {1 - \exp ( - \frac{{\mu A{D^{2}_\blue{u,U-S}}}}{{{\beta _s}}})} \big)^{\alpha _s}}}\left| {D_\blue{u,U-S} = {x}} \right.}\Big]\nonumber\\
&\mathop  =  w\int_{{a_s}}^{{d_{\max }}( {{\theta _c}} )} {\big( {1 -( {\sum\limits_{j = 0}^{{\alpha _s}} {( {_j^{{\alpha _s}}} )} {{( { - 1} )}^j}\exp ( - \frac{{j\mu A{x^2}}}{{{\beta _s}}})} )} \big)} {f_{D_\blue{u,U-S}}}(x)\mathrm{d}x\nonumber\\
&\mathop  = \limits^{\left( c \right)} \frac{w}{{{\theta _c}{r_e}{r_s}}}\int_{{a_s}}^{{d_{\max }}( {{\theta _c}})} {\Big( {\sum\limits_{j = 1}^{{\alpha _s}} {\left( {_j^{{\alpha _s}}} \right)} {{\left( { - 1} \right)}^{j + 1}}\exp ( - \frac{{j\mu Ax^2}}{{{\beta _s}}})} \Big)}\nonumber \\
&\qquad\qquad\qquad\qquad\quad\qquad\qquad\quad\times\frac{{{x}}}{{\sqrt {1 - {( {\frac{{r_e^2 + r_s^2 -x^2}}{{2{r_e}{r_s}}}} )^2}} }}\mathrm{d}{x},
\end{align}
\end{footnotesize}
where (a) is obtained by letting $A = \frac{{\tau \sigma _S^2}}{{\blue{p_u}( {\frac{c}{{4\pi {f_s}}}} )^2}}$. (b) results from Proposition \ref{pro 1} and \eqref{app 2}. (c) is derived by using ${f_{\blue{D_{u,U-S}}}}(x)$ in (\ref{US-cdf}).

\bibliographystyle{IEEEtran}
\vspace{-2mm}
\bibliography{main_arxiv_black}

% \end{CJK}
\end{document}